%% file: main.tex
\documentclass[sigplan, nonacm]{acmart}

\usepackage{hyperref}
\usepackage{hyperxmp}

\usepackage{tikz}
\usepackage{xcolor}
\usepackage[normalem]{ulem}
\usepackage{amsmath}
\usepackage{multirow}
\usepackage{diagbox}
\usepackage{makecell}
\usepackage{color, soul}
\usepackage[capitalise]{cleveref}
\usepackage{multicol} 
\usepackage{float}
\usepackage{enumitem}
\usepackage[linesnumbered,ruled,vlined]{algorithm2e}
\SetCommentSty{small}
\usepackage{stmaryrd}  
\usepackage{hhline}    

\newcommand{\frameworkname}{LightStim}

\begin{document}

\title{\frameworkname: A Framework for QEC Protocol Evaluation and Prototyping with Automated DEM Construction}
\titlenote{This work was primarily led and supported by the U.S. Department of Energy, Office of Science, National Quantum Information Sciences Research Centers, Quantum Science Center. This work was also supported in part by NSF Grants 2534655, 2552449, 2547483, and 2422169.}

\author{
Xiang Fang\textsuperscript{1}, Ming Wang\textsuperscript{2}, Yue Wu\textsuperscript{3}, Sharanya Prabhu\textsuperscript{1}, Dean Tullsen\textsuperscript{1}, Narasinga Rao Miniskar\textsuperscript{4}, Frank Mueller\textsuperscript{2}, Travis Humble\textsuperscript{4}, Yufei Ding\textsuperscript{1}}

\affiliation{
    \institution{
    \textsuperscript{1}University of California, San Diego, CA, USA \\
    \textsuperscript{2}North Carolina State University, Raleigh, NC, USA \\
    \textsuperscript{3}Yale University, New Haven, CT, USA \\
    \textsuperscript{4}Oak Ridge National Laboratory, Oak Ridge, TN, USA}
    \country{}
}

\input{TexFile/01_Abstract}
\maketitle 

\input{TexFile/02_Intro}

\input{TexFile/03_Background}
\input{TexFile/06_Tech}

\input{TexFile/07_Setup}
\input{TexFile/08_Eval}
\input{TexFile/09_Related_Work}
\input{TexFile/10_Conclusion}

\bibliographystyle{plain}
\bibliography{references}

\appendix
\input{TexFile/04_Motivation}
\input{TexFile/11_AppendixB}
\input{TexFile/12_AppendixC}

\end{document}

%% file: TexFile/01_Abstract.tex
\begin{abstract}                                        
  Fault-tolerant quantum computing increasingly demands
  rigorous, circuit-level evaluation of diverse quantum   
  error correction (QEC) protocols and efficient
  prototyping of new ones. Such evaluation requires both  
  the physical circuit and its Detector Error Model
  (DEM) to simulate end-to-end logical error rates. However, DEM construction today is performed by
  manual annotation, a tedious and error-prone process
  that effectively limits evaluation to simple memory
  experiments. We present \textbf{\frameworkname{}}, a
  framework that automates DEM construction concurrently
  with circuit compilation by maintaining a Pauli tableau
  augmented with measurement records, with no
  protocol-specific input required. We benchmark
  \frameworkname{} across protocols from memory
  experiments to end-to-end distillation circuits;
  cross-validation against public implementations confirms
   exact detector and observable counts and consistent
  logical error rates. Additionally, we
  demonstrate a novel heterogeneous cross-code
  lattice surgery design between surface and punctured
  quantum Reed-Muller codes. These capabilities together
  make \frameworkname{} a unified infrastructure for
  systematic QEC protocol evaluation and exploration.
  \frameworkname{} is open-sourced at
  \url{https://github.com/QuTone/LightStim}.
  \end{abstract}

%% file: TexFile/02_Intro.tex
\section{Introduction}\label{sec:intro}

Quantum computing (QC) is entering the fault-tolerant (FT) era~\cite{shor1996fault, preskill2025beyond}, where quantum error correction (QEC)~\cite{gottesman2002introduction, knill1996threshold} is integrated to protect fragile quantum states from physical noise. In recent years, multiple hardware platforms have successfully demonstrated quantum memory with multi-round error correction across diverse code families~\cite{google2023suppressing, acharya2024quantumerrorcorrectionsurface, bluvstein2024logical, zhao2022realization, bluvstein2025architectural, wang2026demonstration, reichardt2411logical, mayer2024benchmarking}. Building on these successes, the community is increasingly turning toward the next frontier: implementing logical operation primitives~\cite{kang2023quantum, postler2022demonstration, rodriguez2024experimental, besedin2025realizing, daguerre2025experimental, postler2024demonstration, dasu2025breaking, kim2024magic} and composing them into larger-scale logical circuits for practical quantum algorithms~\cite{cao2019quantum, daley2022practical, bauer2020quantum, pearson2020simulating, liu2024toward, alexeev2025perspective}.

This progression from memory to computation brings an explosion of QEC protocols: diverse code families (e.g., surface~\cite{fowler2012surface, bravyi1998quantum}, color~\cite{thomsen2024low, gidney2023new}, and qLDPC codes~\cite{gottesman2013fault, tillich2013quantum, bravyi2024high, malcolm2025computing}), multiple computing paradigms (gate-based~\cite{zhou2025low}, Pauli-based~\cite{litinski2019game}, measurement-based~\cite{hastings2021dynamically}), and various realizations of the same logical operation (e.g., transversal gates~\cite{cain2025fast, bluvstein2024logical} and lattice surgery~\cite{horsman2012surface, fowler2018low}, magic state preparation~\cite{litinski2019magic, gidney2024magic} and gate teleportation~\cite{gottesman1999demonstrating}). Evaluating the performance of these diverse protocols and conducting fair comparisons across them is critical for guiding both near-term experiments and long-term architectural decisions.

\begin{figure*}[!ht]
    \centering
    \includegraphics[width=0.99\textwidth]{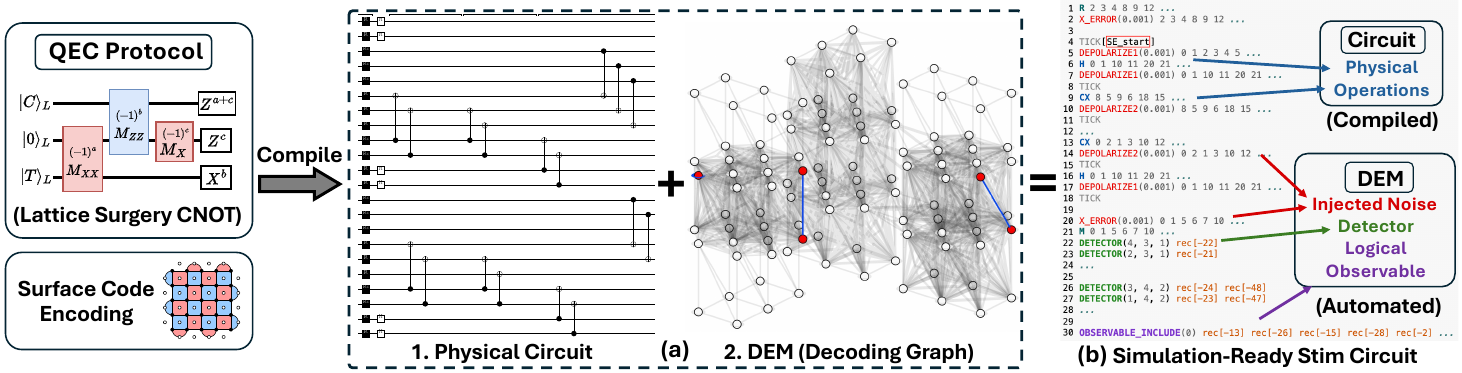}
    \caption{(a) Dual burden on QEC protocol Compilation: Physical Circuit \& DEM. (b) Sample Stim circuit. Physical operations specified by the protocol and the rest are all automated in \frameworkname{}.}
    \label{fig: Intro}
\end{figure*}

The gold standard is to use the \textit{logical error rate} (LER) as the end-to-end metric for evaluation~\cite{dennis2002topological, fowler2012surface, acharya2024quantumerrorcorrectionsurface}, either through physical hardware demonstrations or through \textit{circuit-level noisy simulation} in Stim~\cite{gidney2021stim}. This imposes a \textbf{dual burden} of compilation on the protocol, as shown in Fig.~\ref{fig: Intro}. \textbf{First}, the QEC protocol must be translated into physical operations (gates, measurements, and resets). \textbf{Second}, a \textit{Detector Error Model} (DEM), or \textit{decoding graph}, must be constructed. It represents \textit{detectors} as vertices and the \textit{physical errors} that flip them as edges. The quantum decoder~\cite{higgott2024practical, higgott2022pymatching, wu2025minimum} uses the DEM to infer correction actions and yield the final LER.

In current practice, the DEM is manually annotated in the Stim circuit (Fig.~\ref{fig: Intro}), with \textit{detectors} and \textit{logical observables} as its two key components. However, as we move from basic memory to complex protocols, manual DEM construction becomes infeasible due to three compounding challenges:

\noindent\textbf{(1) Diverse computing mechanisms.}
The same logical operation can have distinct realizations and result in drastically different DEM structures. For example, CNOT via \textit{Lattice surgery} (LS)~\cite{horsman2012surface, fowler2018low, litinski2019game} dynamically creates and destroys auxiliary code patches to mediate interactions, causing the valid detectors to continuously change over time. Conversely, CNOT via \textit{transversal gates} (TG) require \textit{correlated decoding} techniques~\cite{zhou2025low, zhou2024algorithmic, cain2025fast, cain2024correlated}, constructing detectors spatially spanning multiple code blocks. Static, protocol-specific annotations cannot generalize across these different mechanisms.

\noindent\textbf{(2) Explosion of state combinations.}
Unlike memory experiments that focus on a single code block in a fixed state, logical operations involve multiple logical qubits with varying initial states and final measurement bases. Because each combination yields a distinct DEM, the number of possible configurations grows exponentially with the logical qubit count, making any hardcoded approach untenable.

\noindent\textbf{(3) Evolving logical information.}
As operations execute, the logical information evolves through logical gates~\cite{breuckmann2024fold, chen2024transversal}, logical measurements~\cite{cowtan2025parallel, cowtan2024ssip, cowtan2024css}, and adaptive Pauli corrections from feed-forward~\cite{litinski2019game, litinski2019magic}. Manually tracking these transformations is highly error-prone and unscalable; any mistake will provide incorrect information to the decoder and invalidate the LER evaluation.


However, this tedious task can in fact be \emph{fully automated} with correctness guarantees. Our core insights are twofold: (1) QEC protocols admit a unified description via \textit{Pauli string tableau evolution}~\cite{gottesman1997stabilizer, aaronson2004improved}; and (2) DEM construction reduces to \textit{finding deterministic sums of measurement records}, which is equivalent to \textit{finding dependencies among Pauli strings}. These insights motivate our solution: 

\noindent \textbf{First}, augment each tableau Pauli string with its associated measurement records;

\noindent \textbf{Second}, for each new measurement introduced by the physical circuit, recover its corresponding Pauli string via back-propagation of Clifford conjugation; 

\noindent \textbf{Third}, decompose this Pauli string into existing tableau rows (mid-circuit case), or decompose tableau rows into single-qubit readout Paulis (end-circuit case). Each decomposition exposes a Pauli dependency that yields a deterministic sum of measurement records, giving a detector or logical observable.

This solution leads to \textbf{\frameworkname{}}, a general framework that fully automates DEM compilation in lockstep with the physical circuit construction. It accommodates diverse QEC protocols in a unified manner, where even advanced correlated decoding techniques~\cite{cain2024correlated} emerge as a natural subcase. Just as Automatic Differentiation~\cite{baydin2018automatic, paszke2019pytorch} removed gradient derivation as the complexity ceiling on neural network architectures, \frameworkname{} removes DEM complexity as the ceiling on QEC protocols: once the physical circuit is specified, the DEM follows automatically, enabling evaluation of protocols of various structure.

We demonstrate \frameworkname{} across a full spectrum of QEC protocols, ranging from quantum memory to complex logical circuits and novel designs, as summarized in Table~\ref{tab:protocol_summary}. We unify these protocols within a single codebase, rigorously validate our LERs against published literature and theoretical predictions through extensive circuit-level simulations, and uncover previously inaccessible architectural insights into QEC protocol behavior (Sec.~\ref{sec:evaluation}). 


In summary, this paper makes the following contributions:
\begin{itemize}
    \item \textbf{Automated DEM Compilation.} We present a protocol-agnostic algorithm to automatically derive detectors and observables, eliminating the most labor-intensive and error-prone step in QEC protocol evaluation.

    \item \textbf{Systematic QEC Evaluation Framework.} \frameworkname{} provides a modular, configurable framework enabling rapid QEC protocol implementation and rigorous, fair cross-protocol comparison.

    \item \textbf{Open-Source Availability.} We provide the first unified open-source codebase covering the broadest collection of QEC protocols to date.

    \item \textbf{Uncovering Architectural Insights.} We show that gate-specific LERs exceed memory baselines by up to 11.6$\times$, that LS routing distance contributes linearly to teleportation LER, and other findings that directly inform near-term FTQC architecture decisions.
\end{itemize}

%% file: TexFile/03_Background.tex
\section{Background}
\label{sec:background}

This section reviews the FTQC basics (Sec.~\ref{sec:bg:qec}), introduces the Pauli tableau used to track state evolution (Sec.~\ref{sec:bg:tableau}), and explains the construction of DEM needed for decoding and LER estimation (Sec.~\ref{sec:bg:dem}).

\subsection{FTQC: From Memory to Computation}
\label{sec:bg:qec}

FTQC achieves error resilience by encoding logical information into redundant physical qubits~\cite{shor1996fault, gottesman1997stabilizer}, relying on continuous stabilizer measurements to extract error syndromes and a classical \textit{decoder} to correct errors in real time~\cite{gottesman2002introduction, higgott2024practical}.

\vspace{2pt}
\noindent\textbf{From Memory to Computation.} A first milestone is demonstrating \emph{quantum memory}, \textit{preserving} a static encoded state, which has been achieved across diverse codes~\cite{acharya2024quantumerrorcorrectionsurface, reichardt2411logical, bluvstein2024logical, lacroix2025scaling}. The ultimate goal, however, is to \textit{manipulate} encoded information via fault-tolerant logical operations~\cite{gottesman2022opportunities}. Primitive operations have recently been demonstrated on hardware~\cite{kang2023quantum, postler2022demonstration, rodriguez2024experimental, besedin2025realizing, daguerre2025experimental, postler2024demonstration, dasu2025breaking, kim2024magic}, and numerous protocols continue to emerge with increasing complexity~\cite{cowtan2025parallel, yoder2025tour, koh2026entangling}. The community has not yet converged on a definitive architecture; therefore, the ability to rigorously benchmark existing schemes and prototype new designs is vital to navigate this expanding FTQC design space.

\subsection{Pauli Tableau and Update Rules}
\label{sec:bg:tableau}

\begin{figure}[!ht]
    \centering
    \vspace{-6pt}
    \includegraphics[width=0.46\textwidth]{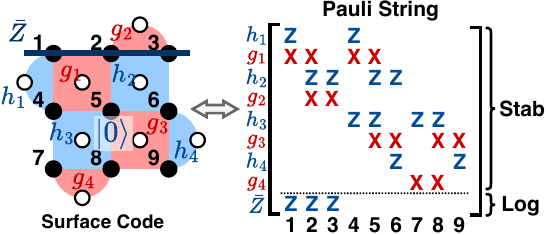}
    \caption{Pauli tableau representation of QEC systems. Stab: stabilizers; Log: logical operators.}
    \label{fig: BG-Tableau}
    \vspace{-6pt}
\end{figure}

\noindent\textbf{Pauli Tableau.} 
QEC relies on stabilizer codes~\cite{gottesman1997stabilizer}, which can be represented by a \textit{Pauli tableau}~\cite{aaronson2004improved}, a matrix where each row represents a Pauli string. As Fig.~\ref{fig: BG-Tableau} shows, a tableau consists of (1) \textit{stabilizer rows} that are measured to detect errors, and (2) \textit{logical operator rows} that define the underlying logical states, e.g., $\bar{Z}$ ($\bar{X}$) indicates a logical $|0\rangle_L$ ($|+\rangle_L$) state. Crucially, this data structure can scale from a single code block to multi-block systems.

\noindent\textbf{Update Rules.} Beyond describing static structures, the Pauli tableau can track dynamic state evolution in Clifford circuits:

\noindent \textbf{Rule 1: Clifford Gates.} A Clifford gate $C$~\cite{gottesman1997stabilizer} maps any Pauli string $P$ to another Pauli string $C P C^\dagger$ via conjugation.

\noindent\textbf{Rule 2: Pauli Measurements.} Measuring a Pauli operator $P$ against the current tableau has two cases:

\noindent \emph{Case 1. Anti-commuting}: If $P$ anti-commutes with any current Pauli rows, choose a pivot row and replace it by $P$, then multiply this pivot row to other anti-commuting rows. The measurement outcome is random $\pm 1$ with 50/50 probability.

\noindent \emph{Case 2. Commuting}: If $P$ commutes with all Pauli rows, it can be expressed as a product of existing Paulis, yielding a deterministic measurement outcome. 

\noindent QEC protocols consist of Clifford circuits augmented with resource states (e.g., magic states~\cite{bravyi2012magic}), so the Pauli tableau serves as a universal abstraction for tracking FTQC system behaviors. For more details, please refer to~\cite{gottesman1998heisenberg, aaronson2004improved}.

\subsection{Detector Error Model and Noisy Simulation}
\label{sec:bg:dem}

\begin{figure}[!ht]
    \centering
    \vspace{-6pt}
    \includegraphics[width=0.46\textwidth]{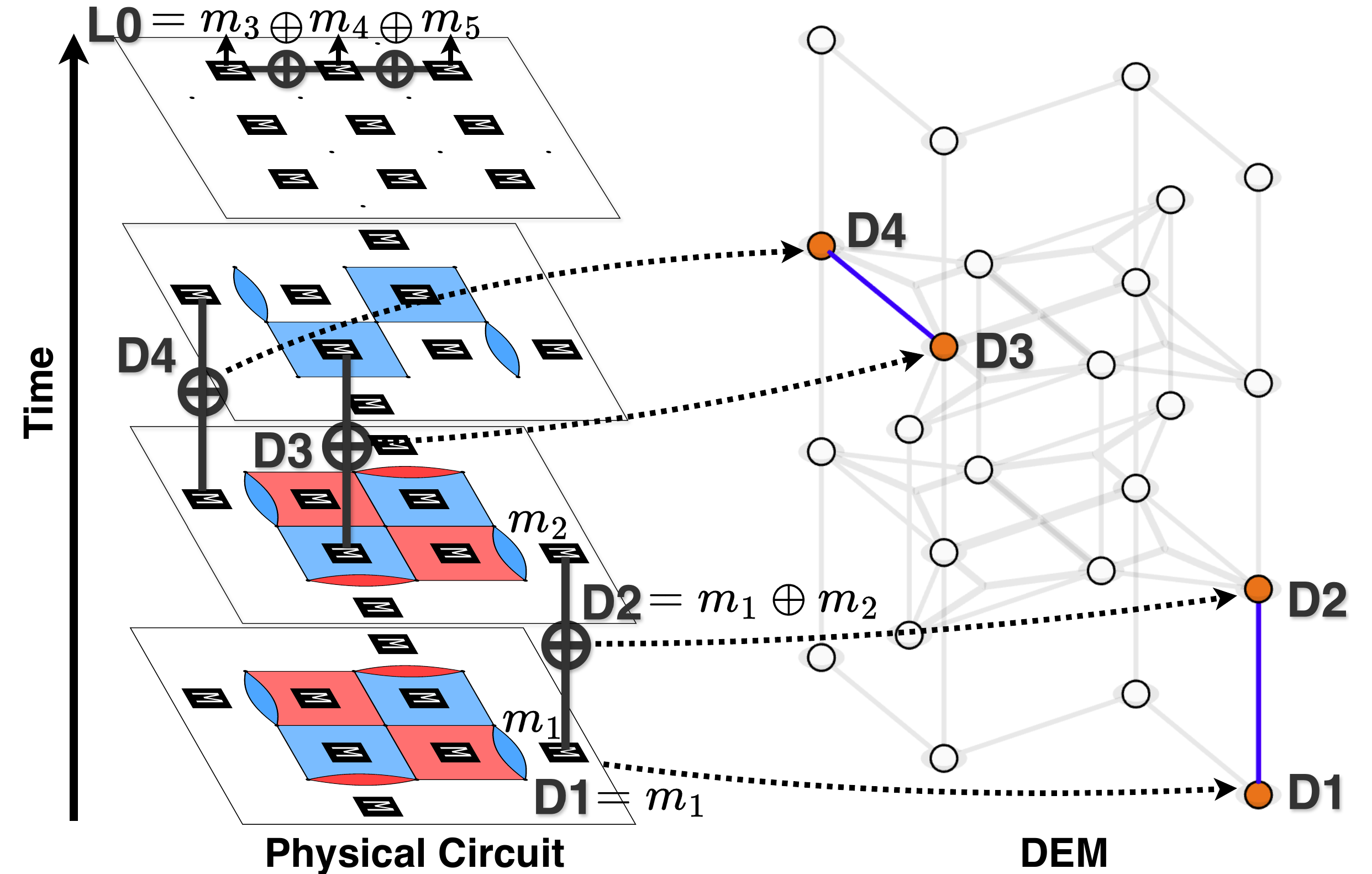}
    \caption{Physical Circuit and DEM construction for surface code Z memory experiment.}
    \label{fig: DEM-BG}
    \vspace{-6pt}
\end{figure}

A FTQC system requires compiling both the \textit{physical circuits} and the \textit{DEM} (Sec.~\ref{sec:intro}). Using the surface code $Z$-memory experiment~\cite{github_stim} as a concrete example, Fig.~\ref{fig: DEM-BG} illustrates the DEM's three core components.

\begin{figure*}[!ht]
    \centering
    \includegraphics[width=0.99\textwidth]{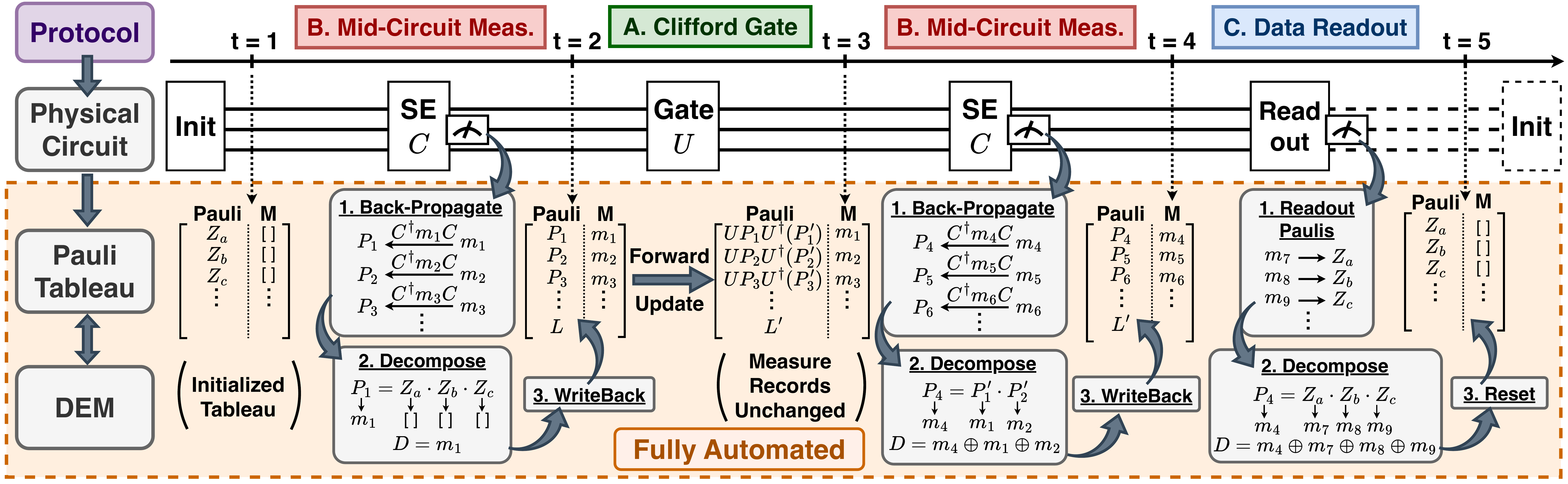}
    \caption{Overview of \frameworkname{}'s Pauli Tracker workflow. The physical circuit drives a forward update of the record-augmented Pauli tableau, triggering DEM construction upon encountering measurements.}
    \label{fig: overview}
\end{figure*}

\vspace{2pt}
\noindent\textbf{(1) Detector: Vertices.}
A detector consists of measurement records with a \textit{deterministic parity} (XOR sum) under noiseless execution, forming the DEM's \textbf{vertices}. In the memory example, a syndrome qubit keeps measuring the same stabilizer; without errors, consecutive measurements must match, yielding a deterministic parity detector (e.g., $D_2 = m_1 \oplus m_{2}$). These detectors encode the information needed to infer where and what errors occurred in the circuit.

\noindent\textbf{(2) Logical Observable: Targets.}
A logical observable is also a deterministic parity of measurement records under noiseless execution, but it is evaluated at the end of execution to verify the final logical state. For instance, the final measurements of data qubits $m_3,m_4,m_5$ form the logical $\bar{Z}$ operator, yielding a deterministic sum corresponding to the state $|0\rangle_L$ and giving the logical observable $L_0 = m_3\oplus m_4\oplus  m_5$. These logical observables represent the encoded information that QEC aims to protect, and they remain unchanged if the decoder successfully identifies and corrects any errors.

\noindent\textbf{(3) Error Mechanism: Edges.}
Error mechanisms are the possible error sources in a circuit given a noise model. In the DEM, each mechanism is represented as an \textit{edge} connecting the detector/observable vertices it would flip. If a mechanism flips more than two vertices, the edge becomes a \textit{hyperedge}. For example, a measurement error in $m_1$ flips two temporal detectors ($D_1, D_2$), and a data qubit $X$ error flips detectors ($D_3, D_4$), forming standard edges. Each edge is weighted by the aggregated probability of its underlying physical errors.

\vspace{2pt}
\noindent\textbf{Noisy Simulation.} Classical noisy simulation for QEC is usually performed using  Stim~\cite{gidney2021stim}, where users inject specific noise models into the physical operations and annotate valid detectors and observables. Stim then derives the corresponding error mechanisms and their probabilities, generating a fully compiled DEM. The simulation then proceeds in a Monte Carlo fashion: each shot corresponds to one experimental run, producing detector events and the true logical observable outcome. The detector events are fed into a decoder (configured with the DEM) to predict a correction, which is compared against the true observable to determine whether the logical state was successfully recovered. After collecting enough shots, the LER can be estimated to compare different protocols and decoders.

%% file: TexFile/06_Tech.tex
\input{TexFile/06_Tech1}

\input{TexFile/06_Tech2}

%% file: TexFile/06_Tech1.tex
\section{\frameworkname{} Core: Pauli Tracker}
\label{sec:tech}


This section illustrates how \frameworkname{}'s core, \textit{Pauli tracker}, constructs the DEM automatically and progressively as the physical circuit is compiled. The tracker maintains a Pauli tableau $\Omega$ where each row is augmented with associated measurement records $M$. 
The tracker performs \emph{three routines} depending on the coming circuit blocks (Fig.~\ref{fig: overview}): 

\noindent\textbf{(A) Clifford Gates.} Pauli strings are updated via conjugation while records remain unchanged (Sec.~\ref{sec:tech:unitary}).

\noindent \textbf{(B) Mid-Circuit Measurements.} Back-propagation recovers the measured Pauli, which is decomposed against current tableau rows to emit detectors; a write-back then refreshes the tableau (Sec.~\ref{sec:tech:mid}). 

\noindent \textbf{(C) Data Readout.} Tableau rows are decomposed into single-qubit readout Paulis to yield final detectors and logical observables, and records are reset (Sec.~\ref{sec:tech:readout}).

\subsection{Clifford Gates: Pauli Tableau Update}
\label{sec:tech:unitary}

When a Clifford gate block (transversal CNOT, fold-transversal H/S, etc.) is appended, the tracker updates $\Omega$ via conjugation (Rule~1, Sec.~\ref{sec:bg:tableau}). This step updates the Pauli strings in $\Omega$ without modifying their associated measurement records (Fig.~\ref{fig: overview}, $t=2$ to $t=3$), ensuring that records always reflect the most up-to-date Pauli strings as the system evolves. Records stay fixed through gate blocks, and their update can only originate from a measurement, regardless of how complex the intervening Clifford circuit is. This gives the first routine.

\subsection{Mid-Circuit Measurements: Detectors}
\label{sec:tech:mid}


Mid-circuit measurements on syndrome qubits trigger back-propagation and Pauli decomposition to emit detectors. This routine drives all syndrome extraction (SE) blocks, whether in memory experiments, in lattice-surgery logical measurements~\cite{litinski2019game, fowler2018low, cowtan2025parallel}, or after transversal gates~\cite{cain2024correlated, zhou2025low}.

\vspace{2pt}
\noindent\textbf{1. Back-propagated Pauli strings.}
Measuring a syndrome qubit $q$ in $Z$ basis after SE block $C$ is equivalent to measuring $\hat{p} = C^{\dagger}Z_q C$ before it. The tracker derives $\hat{p}$ from the physical circuit alone via binary matrix multiplication, without any protocol-specific input, naturally verifying whether the SE block measures the intended Pauli strings. This is especially valuable for non-standard SE designs~\cite{chao2018quantum, gidney2023new}, where manually identifying the actual measured Pauli string can be convoluted and error-prone.

\vspace{2pt}
\noindent\textbf{2. Decomposition and detector construction.}
After back-propagation, $\hat{p}$ is time-aligned to the current $\Omega$ before the newly appended circuit block. Rule~2 (Sec.~\ref{sec:bg:tableau}) dispatches to one of two cases for measuring $\hat{p}$ on $\Omega$:

\noindent\textit{Case A. Anti-commutes, no detector.}
If $\hat{p}$ anti-commutes with any row of $\Omega$, one anti-commuting row is designated the pivot and replaced by $\hat{p}$; all other anti-commuting rows are XOR-updated with the pivot and their measurement records. No detector is emitted in this case. This occurs, for example, at the first SE round after initialization or at every lattice-surgery boundary, where newly introduced stabilizers yield non-deterministic outcomes.

\noindent\textit{Case B. Commutes, detector emitted.}
If $\hat{p}$ commutes with all rows, it decomposes as a product of current stabilizers $P_1,\ldots,P_s$ via GF(2) linear algebra (\textsc{RREF\_Solve}). Since $\hat{p}$ and its Pauli decomposition represent the same Pauli measurement, their records must agree under noiseless execution, yielding a deterministic detector (e.g., Fig.~\ref{fig: overview}, $t=1$ to $t=2$ and $t=3$ to $t=4$):
\begin{equation*}
    D = m_q\oplus \left(m_1\oplus\cdots\oplus m_s\right)
\end{equation*}
Critically, this Case A/B dispatch requires no protocol-specific knowledge. This generality enables \frameworkname{} to discover detectors that are absent in manually annotated circuits, such as the additional detectors arising from linearly dependent stabilizers in BB codes and toric codes.

\noindent\textbf{3. Write-back and canonical basis.}
After each SE block, we re-express $\Omega$ in the freshly measured Paulis as the new canonical basis (\textsc{WriteBack}): tableau rows that depend on the new basis are rewritten accordingly, while independent rows are identified as logical operators. This automatic classification tracks the evolution of logical information and captures the progressive reduction of logical operators under logical measurement~\cite{vuillot2019code}. In addition, the \textsc{WriteBack} design aligns the tableau to the freshly measured records after every SE block, so subsequent detector decompositions reference only the most recent layer of measurement records, yielding temporally local detectors. Without it, decompositions could span arbitrarily many past rounds, producing high-degree hyperedges that complicate the DEM structure. Alg.~\ref{alg:mid} summarizes the full mid-circuit measurement routine.

\begin{algorithm}
\caption{ProcessMidMeasurement($\hat{p}$, $i$)}
\label{alg:mid}
\KwIn{Pauli$_\textbf{Back-Propagated}$ $\hat{p}$\,; current record index $i$}
\tcp{buf[$i'$] stores (back-propagated Pauli, record index) for syndrome meas. $i'$}
$A \leftarrow \{j : \Omega[j] \text{ anti-commutes with } \hat{p}\}$\;
\eIf{$A \neq \emptyset$}{
    \tcp{Case A: anti-commutes, no detector}
    pivot $\leftarrow A[0]$\;
    \lFor{$j \in A \setminus \{\mathrm{pivot}\}$}{$\Omega[j] \leftarrow \Omega[j] \oplus \Omega[\mathrm{pivot}]$}
    $\Omega[\mathrm{pivot}] \leftarrow \hat{p}$\;
}{
    \tcp{Case B: commutes, emit detector}
    $\mathbf{c} \leftarrow \textsc{RREF\_Solve}(\Omega,\;\hat{p})$ \tcp*{over GF(2)}
    \textbf{emit} \texttt{DETECTOR}$\!\left(\{\mathrm{buf}[i'] : c_{i'}{=}1\} \cup \{\mathrm{rec}(i)\}\right)$\;
}
$\mathrm{buf}[i] \leftarrow (\hat{p},\;i)$\;
\textsc{WriteBack}$(\Omega)$\;
\end{algorithm}

\vspace{-8pt}
\subsection{Data Readout: Detectors and Logical Observables}
\label{sec:tech:readout}

Data qubit measurements collapse logical states. This occurs when a QEC patch is recycled mid-circuit or when all qubits are measured at the end. Let $\{f_j\}$ denote the single-qubit readout Paulis (e.g., $Z_j$ for computational-basis readout), with base record index $i_0$ (the most recent record). We first update the tableau using these Pauli measurements, then decompose remaining rows of $\Omega$ into $\{f_j\}$: stabilizer rows yield final detectors, and logical rows yield logical observables.

\noindent\textbf{1. Tableau update by Readout Paulis.} The tracker first updates $\Omega$ using $\{f_j\}$ via Rule~2 (Sec.~\ref{sec:bg:tableau}), XOR-updating any rows that anti-commute with the readout Paulis.

\begin{algorithm}
\caption{ProcessDataReadout($\{f_j\}$, $i_0$)}
\label{alg:readout}
\KwIn{readout Paulis $\{f_j\}$\,; base record index $i_0$}
\tcp{Pass 1: final detectors from stabilizer rows}
\For{each stabilizer row $s \in \Omega$}{
    $\mathbf{c} \leftarrow \textsc{RREF\_Solve}(\{f_j\},\;s)$ \tcp*{GF(2);}
    \textbf{emit} \texttt{DETECTOR}$\!\left(\{i_0{+}j : c_j{=}1\} \cup \{\text{recs of }s\}\right)$\;
}
\tcp{Pass 2: logical observables from logical rows}
\For{each logical row $l_k \in \Omega$}{
    $\mathbf{c} \leftarrow \textsc{RREF\_Solve}(\{f_j\},\;l_k)$ \tcp*{GF(2);}
    \textbf{emit} \texttt{OBSERVABLE\_INCLUDE}$(k,\;\{i_0{+}j : c_j{=}1\})$\;
}
\textsc{Reset}$(\Omega)$\;
\end{algorithm}

\noindent\textbf{2. Decomposition and DEM construction.}

\noindent\textit{Pass 1. Final detectors.} Each remaining stabilizer row $s_i \in \Omega$ decomposes into a linear combination of $\{f_j\}$ via \textsc{RREF\_Solve}. The intuition is: the measurement records of $s_i$ and the product of its constituent data-qubit measurements must agree under noiseless execution, forming a deterministic detector.

\noindent\textit{Pass 2. Logical observables.} Each logical row $l_k \in \Omega$ decomposes into $\{f_j\}$ similarly, XOR-ing the associated records to extract the logical value. In general protocols, logical observables may span both data and syndrome records, encoding feed-forward Pauli corrections, which \frameworkname{} tracks automatically. Alg.~\ref{alg:readout} summarizes both passes.

\noindent\textbf{3. Reset.} Recycled readout qubits have their measurement records cleared to prevent stale records from corrupting future detector construction. This allows \frameworkname{} to accommodate repeated qubit recycling and re-initialization, and more broadly, dynamic resource allocation throughout a protocol's lifespan.

\noindent Together, the above three routines (Sec.~\ref{sec:tech:unitary}-Sec.~\ref{sec:tech:readout}) form a protocol-agnostic engine for automated DEM compilation, applying to any QEC protocol and any combination of initial states, gates, and readout bases, with no manual annotation required. Appendix~\ref{sec:motivation} walks through two-qubit logical teleportation in both lattice surgery (LS) and transversal gates (TG), showing how \frameworkname{} naturally derives correlated decoding (TG) and feed-forward corrections (LS).

%% file: TexFile/06_Tech2.tex
\section{\frameworkname's Pipeline for FTQC Evaluation}
\label{sec:tech2}

Building on the core tracker (Sec.~\ref{sec:tech}), \frameworkname{} extends it into a modular, reusable pipeline for end-to-end QEC protocol evaluation. The pipeline consists of three key components: a QEC system abstraction that defines the spatial-temporal representation (Sec.~\ref{sec:tech2:QECSystem}), atomic operations that compose general protocols (Sec.~\ref{sec:tech2:ops}), and a simulation backend for standardized noise injection and decoding (Sec.~\ref{sec:tech2:sim}).

\subsection{QEC System Assembly}
\label{sec:tech2:QECSystem}

\noindent\textbf{QEC patch.} \frameworkname{} provides a library of major QEC code families (repetition, surface, toric, BB, PQRM, etc.), each packaged as a self-contained \emph{QEC Patch} with qubit coordinates, stabilizers, and logical operators. SE circuits and logical operations (e.g., transversal gates) are defined as independent, reusable modules within each family, enabling modular composition across protocols.

\noindent\textbf{QEC system.} Patches are placed into a \emph{QEC system} in a define-by-run manner: they can be added at any point during protocol composition with the Pauli tableau extending adaptively, mirroring dynamic resource allocation on real hardware. Each patch's position and orientation are specified on a shared qubit plane to reflect the target architectural layout. The system tracks qubit coordinates and active stabilizers as they evolve in time, serving as the spatial-temporal substrate for protocol construction and readily extending to heterogeneous multi-code systems.

\subsection{Atomic Operations Composition}
\label{sec:tech2:ops}

We define five \emph{atomic operations} that are highly expressible for composing general QEC protocols. Each corresponds to a routine in our Pauli Tracker (Sec.~\ref{sec:tech}) or a status update of the QEC system, enabling concurrent DEM construction as the protocol is composed.

\noindent\textbf{1. Initialize}($\mathit{basis}$): initialize data qubits in the specified basis and register their Pauli strings in the tracker's tableau.

\noindent\textbf{2. SyndromeExtraction}($r$): run $r$ SE rounds; the first invokes the mid-measurement routine (Sec.~\ref{sec:tech:mid}), and subsequent rounds compile into a repeated block with pairwise detectors, eliminating redundant tableau decompositions.


\noindent\textbf{3. Activate\,/\,DeactivateCoupler}: introduce or remove a \emph{logical coupler}: an ancilla patch connecting multiple QEC patches for lattice surgery. It updates the active stabilizer set of the QEC system (Sec.~\ref{sec:tech2:QECSystem}), so a single SE block configuration dynamically retargets to different regions of the system. Sequential lattice surgery operations thus reduce to toggling couplers on and off.

\noindent\textbf{4. UnitaryBlock}(\textit{gates}): apply Clifford gates, updating the tableau via conjugation without emitting detectors (Sec.~\ref{sec:tech:unitary}).

\noindent\textbf{5. DataReadout}($\mathit{basis}$): measure all data qubits and emit final detectors and logical observables (Sec.~\ref{sec:tech:readout}).

Representative protocols reduce to short compositions of these five operations, for example:

\noindent\emph{Memory Experiment.} Init, SE($d$),Readout; 

\noindent\emph{Transversal Gate.} Init, SE($d$), Unitary,  SE($1$), Readout; 

\noindent\emph{LS CNOT.} Init, SE($d$), (ActCoup, SE($d$), DeactCoup)$\times 2$, Readout; General circuits compose these primitives further.

\vspace{2pt}
\noindent\textbf{Repeat-block optimization.}
SyndromeExtraction($r$) invokes the full mid-measurement routine only for round~1. After \textsc{WriteBack} (Sec.~\ref{sec:tech:mid}), $\Omega$ is re-expressed in the freshly measured Paulis, so rounds $2,\ldots,r$ back-propagate identical Pauli strings and produce the same Case-B decompositions---each detector reduces to a pairwise XOR of consecutive syndrome records. \frameworkname{} exploits this fixed-point property to compile rounds $2,\ldots,r$ into a single Stim \texttt{REPEAT} block, replacing $O(r)$ RREF decompositions with one. The resulting $O(r)$ reduction in compilation cost directly explains why large-scale experiments (e.g., surface code at $d{=}31$ with 29{,}761 detectors (Table~\ref{tab:compilation}) compile in just seconds.

\vspace{-2pt}
\subsection{Noise Injection and Simulation}
\label{sec:tech2:sim}

\noindent\textbf{Standardized noise injection.} We introduce a \emph{Noise Injector} that standardizes the noise injection process across all protocols. Unlike existing tools that require per-operation noise annotation, \frameworkname{} automatically injects noise after the clean circuit is constructed, based on a unified noise specification. We support code-capacity, phenomenological, circuit-level, $XZ$-biased, and general configurable Pauli channels~\cite{higgott2024practical}, enabling fair cross-protocol comparison under identical noise conditions.

\begin{figure}[!ht]
    \centering
    \includegraphics[width=0.48\textwidth]{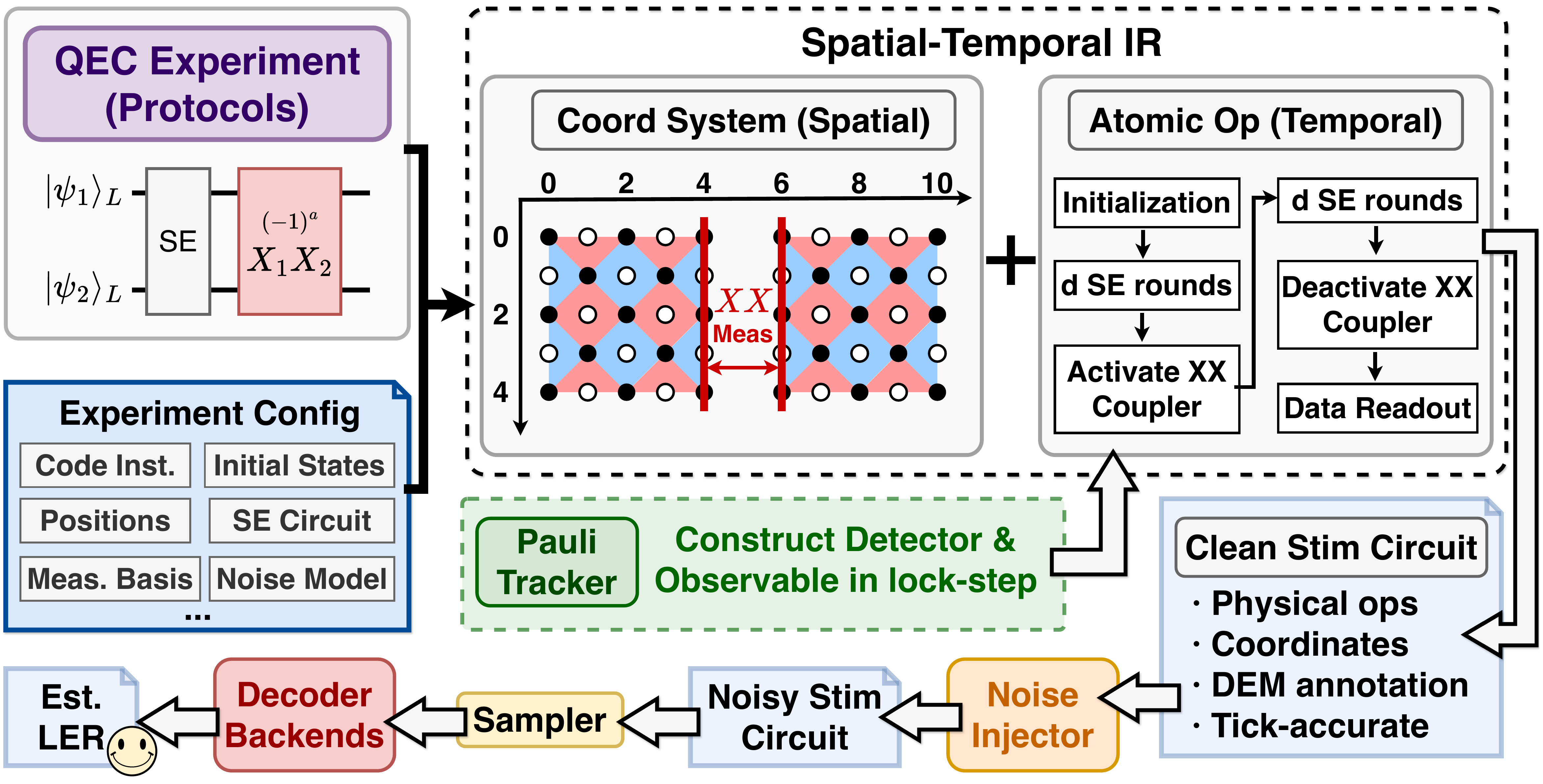}
    \caption{Complete workflow of \frameworkname{} from protocol implementation to evaluation results.}
    \vspace{-4pt}
    \label{fig: workflow}
    \vspace{-6pt}
\end{figure}

\noindent\textbf{Decoder backend.} The noisy circuit is processed by Stim~\cite{gidney2021stim}, which generates samples from the DEM, passes them through a unified post-selection stage if applicable (e.g., state injection~\cite{li2015magic, lao2022magic}, magic state distillation~\cite{litinski2019magic}), and feeds into a decoder backend: CPU decoders (PyMatching~\cite{higgott2022pymatching}, BPOSD~\cite{wang2025fully, nvidiaCUDAQx2025, roffe2020decoding, panteleev2021degenerate}, MWPF~\cite{wu2025minimum}) and GPU BPOSD via CUDA-Q~\cite{nvidiaCUDAQx2025}. This collection handles both simple graph-like DEMs via fast matching and complex DEMs with hyperedges.

This unified simulation pipeline provides the foundation for rigorous, fair comparison across QEC protocols under a standardized noise profile and decoder configuration. Crucially, decoupling circuit construction from noise injection means the same protocol implementation can be evaluated under any noise model without code duplication, a capability that existing Stim-based workflows do not provide. Quantitative compilation performance and productivity measurements across all protocols are reported in Sec.~\ref{sec:eval:compilation}.

\begin{table*}[t]
\centering
\caption{QEC protocol evaluation benchmarks categorized by complexity. \frameworkname{} unifies evaluation across diverse code families and protocols within a single framework.}
\label{tab:protocol_summary}
\small
\begin{tabular}{@{} l p{4.7cm} p{4.4cm} p{4.7cm} @{}}
\toprule
\textbf{Category} & \textbf{Protocols \& Codes} & \textbf{Prior Open-Source Availability} & \textbf{\frameworkname{} Provides} \\
\midrule
\textbf{A. Memory} 
& Surface~\cite{fowler2012surface, horsman2012surface}, Toric~\cite{kitaev2003fault}, Color~\cite{gidney2023new}, BB codes~\cite{bravyi2024high}, 4D Code~\cite{aasen2025topologically} 
& Fragmented per-code scripts~\cite{gidney2021stim, github_qLDPC, github_chromobius, gong2024toward}; 4D Code N/A 
& Unified codebase across all major QEC code families \\
\midrule
\multirow{3}{*}{\shortstack[l]{\textbf{B. Logical} \\ \textbf{Operations}}} 
& Transversal Gates (H, S, CNOT)~\cite{zhou2025low, breuckmann2024fold, chen2024transversal} 
& Hardcoded circuit examples for fixed $d$ without source codes~\cite{zhou2025low} 
& Parameterized circuits with automated correlated decoding \\
\cmidrule{2-4}
& Lattice Surgery (ZZ, XX, CNOT, Pauli-product measurement)~\cite{litinski2019game, horsman2012surface} 
& Hardcoded two-patch examples for fixed $d$~\cite{stackexchange_lattice_surgery} 
& Multi-patch couplers with flexible patch positions\\
\cmidrule{2-4}
& State Injection (Middle, Corner)~\cite{li2015magic, lao2022magic} 
& Separate pipelines per scheme~\cite{yin2026iswitch} 
& Unified injection with configurable post-selection targets\\
\midrule
\multirow{2}{*}{\shortstack[l]{\textbf{C. Logical} \\ \textbf{Circuits}}} 
& Bell Teleportation~\cite{gottesman1999demonstrating} 
& Closed-source only~\cite{stack2025assessing} 
& First open-source TG \& LS implementation in a unfied codebase\\
\cmidrule{2-4}
& Magic State Distillation~\cite{zhou2025low, litinski2019game} 
& TG: hardcoded~\cite{zhou2025low}; LS: none 
& First open-source LS implementation; parameterized TG \\
\midrule
\textbf{D. New Protocol} 
& Cross-code LS (Surface $\leftrightarrow$ PQRM) 
& None (novel) 
& First heterogeneous-code implementation with automated DEM \\
\bottomrule
\end{tabular}
\end{table*}

\subsection{Workflow Summary}
\label{sec:tech2:workflow}

Fig.~\ref{fig: workflow} summarizes the end-to-end \frameworkname{} workflow. The user specifies a QEC experiment via a protocol (e.g., memory, transversal gate, lattice surgery) and an \emph{experiment configuration} (e.g., code instance, qubit positions, SE circuit, measurement basis, initial states, noise model). \frameworkname{} compiles this into a \emph{Spatial-Temporal IR}: the spatial component places all patches on a shared coordinate plane via \textit{QEC system} description (Sec.~\ref{sec:tech2:QECSystem}), while the temporal component orchestrates the \textit{atomic operations} over time (Sec.~\ref{sec:tech2:ops}). As the IR is constructed, the \textit{Pauli Tracker} builds detectors and logical observables in lock-step with the physical operations, rather than as a separate post-processing pass. The resulting tick-accurate clean Stim circuit then enters the simulation backend: the \textit{Noise Injector} applies a chosen noise model, the noisy circuit is sampled, and the events are routed through the unified \textit{decoder backend} to yield the estimated logical error rate. 

Crucially, the IR reduces a QEC protocol to a concise composition over the spatial layout and atomic operations, while the remaining stages are largely automated. The framework is further organized along three axes: \emph{modularity}, as QEC patches, SE circuits, logical operations, noise models, and decoders are independently encapsulated components; \emph{reusability}, as each component plugs into any new protocol once integrated; and \emph{extensibility}, as new codes, gates, noise channels, or decoders enter the framework without touching existing protocols. Building a new experiment thus reduces to specifying a configuration over these building blocks, making \frameworkname{} a unified substrate for fair cross-protocol comparison and rapid prototyping of new QEC designs.

%% file: TexFile/07_Setup.tex
\section{Experiment Setup}
\label{sec:setup}

We evaluate \frameworkname{} in two ways: (1) \textbf{validate} its compilation against existing QEC protocols (Sec.~\ref{sec:setup:existing}), 
and (2) \textbf{demonstrate} its ability to rapidly prototype unexplored protocols (Sec.~\ref{sec:setup:new}). Table~\ref{tab:protocol_summary} summarizes these benchmarks and highlights \frameworkname{}'s newly enabled capabilities. Finally, Sec.~\ref{sec:setup:pipeline} details our simulation pipeline and assumptions.

\subsection{Validation: Existing Protocols}
\label{sec:setup:existing}
We benchmark \frameworkname{} across all existing protocols in Categories A–C of Table~\ref{tab:protocol_summary}, which summarizes the specific codes, prior availability, and \frameworkname{}'s added value for each. Prior to our work, these protocols relied on fragmented, code-specific scripts or isolated static Stim circuits such in Category A (memory experiments) and Category B (logical operations); for Category C (complex circuits), open-source implementations were entirely absent. \frameworkname{} composes all circuits from the atomic operations in Sec.~\ref{sec:tech2} and automates DEM construction with no manual annotation.


\begin{figure}[!ht]
    \centering
    \includegraphics[width=0.48\textwidth]{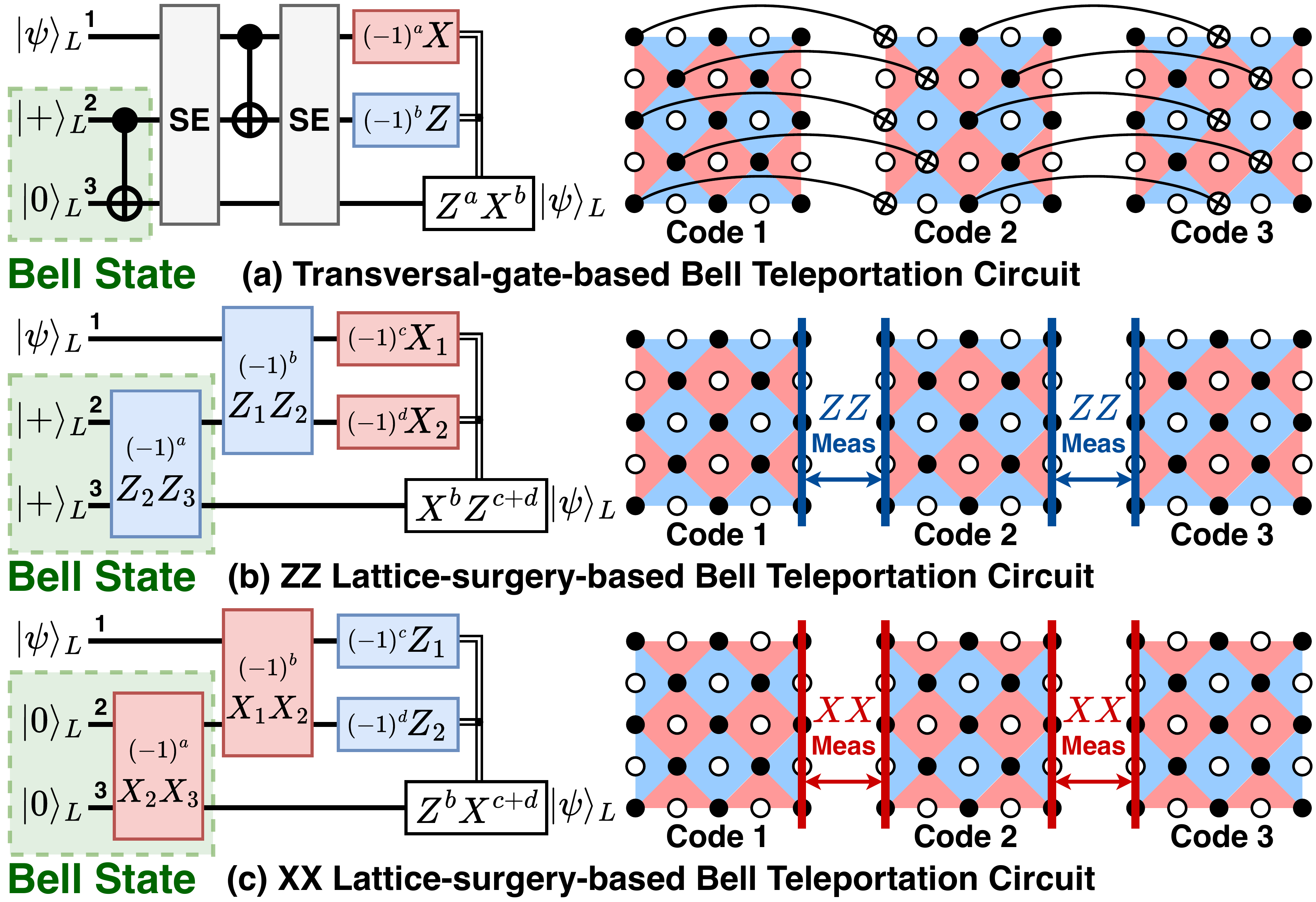}
    \caption{Three realizations of Bell state teleportation circuits: (a) transversal gates, (b,c) $ZZ/XX$ lattice surgery.}
    \label{fig: BellTeleport}
    \vspace{-5pt}
\end{figure}



\subsection{Exploration: Prototyping New Protocols}
\label{sec:setup:new}
To demonstrate \frameworkname{}'s rapid prototyping capability, we prototype and evaluate a novel protocol: \textit{Cross-Code Lattice Surgery (CrossLS)} between surface and punctured quantum Reed-Muller (PQRM) codes~\cite{gong2024computation, steane2002quantum}.

\noindent\textbf{Motivation.} PQRM codes natively support transversal non-Clifford gates~\cite{gong2024computation}. CrossLS leverages this by preparing magic states via transversal gates in a PQRM block and teleporting them into a surface code for reliable storage through LS-$ZZ$ measurement (circuit in Appendix~\ref{sec:motivation}, Fig.~\ref{fig: Teleportation}(a)).

\noindent\textbf{Challenge.} PQRM codes have high-weight stabilizers (scaling as $2^k$) that preclude standard SE methods. Realizing a logical $ZZ$ measurement across heterogeneous code boundaries demands tick-accurate scheduling, and manually deriving the DEM for such merges is intractable.

\begin{figure}[!ht]
    \centering
    \vspace{-8pt}
    \includegraphics[width=0.48\textwidth]{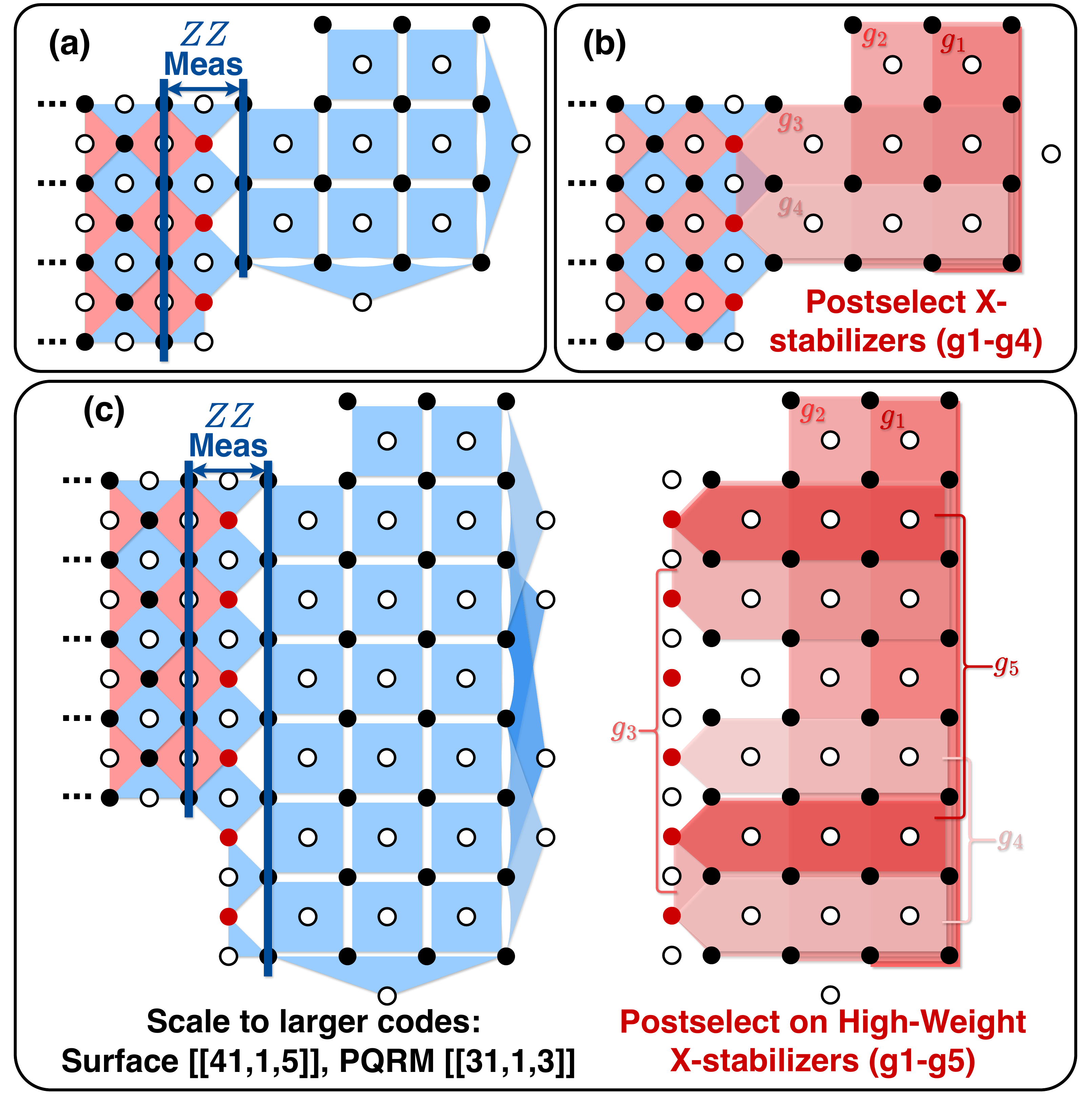}
    \caption{Cross-code Lattice surgery between surface code and punctured quantum Reed-Muller (PQRM) code.}
    \label{fig: magiccross}
    \vspace{-6pt}
\end{figure}

\noindent\textbf{Solution: Hybrid SE-Postselect.} We target a PQRM family (PQRM(1,2,4) $\llbracket 15,1,3 \rrbracket$, PQRM(1,3,5) $\llbracket 31,1,3 \rrbracket$, PQRM(1,4,6) $\llbracket 63,1,3 \rrbracket$) supporting transversal $T$, $T^{1/2}$, $T^{1/4}$ non-Clifford gates, whose $Z$-stabilizers reduce to weight-4, isolating high-weight $X$-stabilizers. We co-locate both codes on a unified square lattice and design a novel 6-tick joint SE circuit that simultaneously measures surface code stabilizers, PQRM $Z$-stabilizers, and the boundary $ZZ$ operator (see Appendix~\ref{sec: tick scheduling}). After $d_{\text{surf}}$ SE rounds, the PQRM patch is measured in $X$-basis for post-selection of high-weight $X$-stabilizer errors (Fig.~\ref{fig: magiccross}). \frameworkname{} automatically tracks the Pauli evolution across this heterogeneous boundary and compiles the end-to-end DEM with no manual annotation.

\subsection{Simulation and Decoding Pipeline}
\label{sec:setup:pipeline}
\noindent\textbf{Decoding Configuration.} We adopt a standard circuit-level noise model~\cite{bravyi2024high, higgott2024practical} for all experiments. For DEMs without hyperedges (e.g., surface- and toric-code memory, LS-based circuits), we employ the PyMatching decoder~\cite{higgott2022pymatching}. For complex DEMs with hyperedges (e.g., BB code memory and TG-based circuits), we utilize BPOSD~\cite{wang2025fully, nvidiaCUDAQx2025, roffe2020decoding, panteleev2021degenerate} (CPU and GPU) and MWPF~\cite{wu2025minimum}. To balance decoding accuracy and latency, we configure BPOSD with \texttt{bp\_iteration} $= 1000$ and \texttt{osd\_order} $= 10$, and MWPF with \texttt{cluster} $= 50$.

\noindent\textbf{Simulation Strategy.} Simulations run on AMD EPYC 9534 CPUs and NVIDIA H100 GPUs. To balance statistical rigor and evaluation throughput, Monte Carlo sampling terminates upon reaching either 100 detected errors or $10^9$ shots, resolving LERs down to $\sim 10^{-7}$.

%% file: TexFile/08_Eval.tex
\section{Evaluation}
\label{sec:evaluation}

We first validate \frameworkname{}'s correctness and compilation efficiency (Sec.~\ref{sec:eval:correctness}--\ref{sec:eval:compilation}), then present simulation results across the existing protocols in Sec.~\ref{sec:setup:existing} and extract FTQC design insights (Sec.~\ref{sec:eval:memory}--\ref{sec:eval:circuits}), and finally evaluate our new CrossLS protocol (Sec.~\ref{sec:eval:crossls}).

\subsection{Correctness Validation}
\label{sec:eval:correctness}

\noindent\textbf{Comparison with open-source references (Table~\ref{tab:correctness}).}
Where open-source implementations exist, we directly compare detector counts and LER.
\emph{(i) Rotated surface code:} at $d\in\{3,5,7\}$, detector and logical observable counts match Stim's built-in reference \emph{exactly}. Under matching circuit-level noise and decoder settings, LER ratios are $1.001\times$, $1.005\times$, and $0.984\times$ for $d{=}3,5,7$ respectively.
\emph{(ii) BB code $Z$-memory:} against reference constructions~\cite{github_sliding_window, maan2026decoding}, \frameworkname{} generates \textit{12 additional detectors} per instance, induced from dependencies among $X$-stabilizers that manual annotations overlook but our Pauli tracker discovers automatically. LER is consistent within $2\times$, attributable to decoder configuration variance.

\vspace{-6pt}
\begin{table}[h]
\centering
\small
\caption{Correctness validation with open-source references.}
\vspace{-5pt}
\label{tab:correctness}
\begin{tabular}{@{}llrrc@{}}
\toprule
Protocol & Instance & Det.\ (Ref.) & Ours & LER ratio \\
\midrule
\multirow{3}{*}{\shortstack[l]{Rotated SC\\(vs.\ Stim)}}
 & $d=3$ & 24  & 24  & $1.001\times$ \\
 & $d=5$ & 120 & 120 & $1.005\times$ \\
 & $d=7$ & 336 & 336 & $0.984\times$ \\
\midrule
\multirow{2}{*}{\shortstack[l]{BB Memory\\(vs.\ pub.~\cite{bravyi2024high, gong2024toward})}}
 & $[[72,12,6]]$   & 432 & 444 & $0.90\times$ \\
 & $[[144,12,12]]$ & 1728 & 1740& $2.0\times$ \\
\bottomrule
\end{tabular}
\end{table}

\vspace{-4pt}
\noindent\textbf{Correctness beyond reference implementations.}
For protocols without open-source references (e.g., lattice surgery, Bell teleportation, distillation), we establish correctness via three cross-checks:
(1)~\frameworkname{} verifies end-to-end logical degree-of-freedom consistency: any mismatch between initialized and measured logical counts raises an error during protocol composition;
(2)~Stim itself rejects non-deterministic detectors at sample time, so any structurally invalid DEM is caught before simulation;
(3)~the resulting LER curves match theoretical predictions (e.g., LER suppression, the $7p^3$ distillation curve in Sec.~\ref{sec:eval:circuits}), providing convergent empirical evidence.
Together, these checks give us high confidence in correctness across all evaluated protocols.

\begin{figure*}[!ht]
    \centering
    \includegraphics[width=0.99\textwidth]{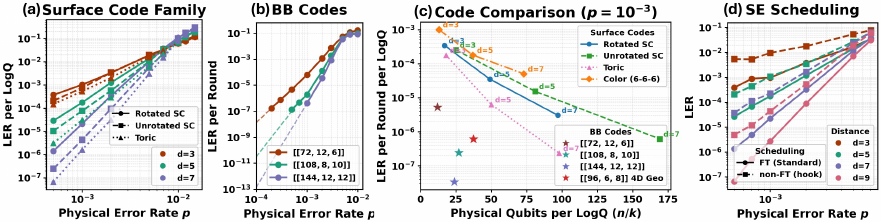}
    \caption{Evaluation of memory experiments. (a) Surface code family. (b) BB code family. (c) QEC efficiency: LER per physical qubit. (d) Effect of SE circuit design on LER.}
    \label{fig: memory}
    \vspace{-8pt}
\end{figure*}

\subsection{Compilation Performance and Productivity}
\label{sec:eval:compilation}

\frameworkname{}'s DEM compilation is a one-time upfront cost per experiment configuration, after which all simulation shots reuse the compiled DEM. Table~\ref{tab:compilation} reports compilation time and productivity across all evaluated protocols.

\begin{table}[!ht]
\centering
\small
\setlength{\tabcolsep}{4pt}
\caption{Compilation time and atomic operation's productivity. Ann.~LoC: annotation instructions (detectors, logical observables) auto-generated by \frameworkname{}; AO~LoC: Atomic operation lines to specify the protocol.}
\label{tab:compilation}
\vspace{-6pt}
\begin{tabular}{@{}lrrrrc@{}}
\hline
\textbf{Protocol} & \textbf{\textit{d}} & \textbf{Qubits} & \textbf{Ann.~LoC} & \textbf{AO~LoC} & \textbf{Time} \\
\hline
\multirow{2}{*}{Surface Mem.}    & 11 &   241 &    1,321 & \multirow{2}{*}{3}  & 234\,ms \\
                                  & 31 & 1,921 &   29,761 &                     &  24.4\,s \\
BB~$[[72,12,6]]$                  &  6 &   144 &      456 & \multirow{2}{*}{3}  & 174\,ms \\
BB~$[[144,12,12]]$                & 12 &   288 &    1,752 &                     & 599\,ms \\
\hline
\multirow{2}{*}{TG CNOT}          &  7 &   338 &    1,346 & \multirow{2}{*}{5}  & 861\,ms \\
                                   & 15 & 1,682 &   13,442 &                     &  52.4\,s \\
\multirow{2}{*}{LS CNOT}          &  7 &   845 &    6,457 & \multirow{2}{*}{6}  &   2.4\,s \\
                                   & 15 & 4,205 &   69,273 &                     & 198\,s \\
\hline
\multirow{2}{*}{Bell Tele.~(TG)}  &  7 &   507 &    2,269 & \multirow{2}{*}{7}  &   2.2\,s \\
                                   & 15 & 2,523 &   21,421 &                     & 363\,s \\
\multirow{2}{*}{Bell Tele.~(LS)}  &  7 &   533 &    5,377 & \multirow{2}{*}{6}  &   1.2\,s \\
                                   & 15 & 2,581 &   57,121 &                     &  79.0\,s \\
\hline
\multirow{3}{*}{TG Distill.}      &  3 &   375 &    1,250 & \multirow{3}{*}{8}  &   1.9\,s \\
                                   &  5 & 1,215 &    5,376 &                     &  76.0\,s \\
                                   &  7 & 2,535 &   13,822 &                     & 753\,s \\
\multirow{3}{*}{LS Distill.}      &  3 &   272 &    1,736 & \multirow{3}{*}{15} & 757\,ms \\
                                   &  5 &   768 &    8,556 &                     &  20.0\,s \\
                                   &  9 & 2,528 &   51,764 &                     & 416\,s \\
\hline
\multirow{2}{*}{CrossLS}          &  3 &    55 &       89 & \multirow{2}{*}{7}  &  32\,ms \\
                                   &  7 &   301 &    1,113 &                     & 584\,ms \\
\hline
\end{tabular}
\end{table}

\noindent\textbf{Productivity.} The AO~LoC column counts the atomic operations (Sec.~\ref{sec:tech2:ops}) needed to specify each protocol; the Ann.~LoC column counts the annotation instructions \frameworkname{} produces automatically. The gap is consistently orders of magnitude: LS CNOT at $d{=}15$ requires 69{,}273 annotations from just 6 atomic operations, and LS distillation produces tens of thousands of annotations from 15 operations. The absence of prior open-source Stim implementations for several benchmarks (e.g., LS distillation) reflects that manual annotation at this complexity is practically infeasible.

\noindent\textbf{Scalability.} Compilation time is dominated by $O(n^3)$ GF(2) row reduction (worst-case) against the $n$-row tableau in the Pauli tracker, where $n$ is the active system size. Table~\ref{tab:compilation} spans five orders of magnitude in compilation time, from milliseconds for small memory experiments to 753\,s for TG distillation at $d{=}7$ (2{,}535 qubits, 13{,}822 detectors across 15 entangled patches)---the most demanding protocol in our suite. At the largest tested scale---thousands of qubits and tens of thousands of detectors---\frameworkname{} remains tractable and covers system sizes relevant to near-term FTQC experiments projected for the coming years. Scaling to substantially larger circuits may benefit from partitioned or sliding-window compilation strategies, which we leave to future work.

\begin{figure*}[!ht]
    \centering
    \includegraphics[width=0.99\textwidth]{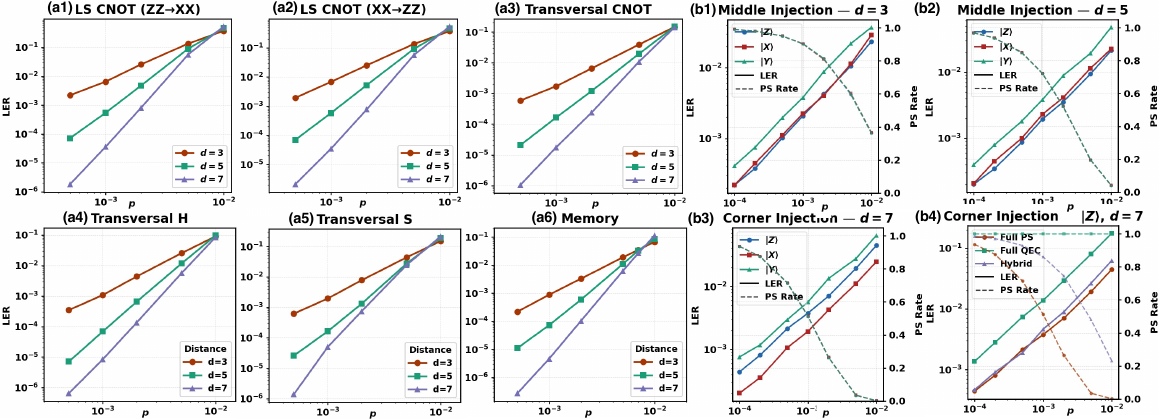}
    \caption{Comprehensive evaluation of logical operations. (a) Transversal gates and lattice surgery operations of unrotated surface code against the memory baseline; (b) State injection of rotated surface code; (b1)-(b3) Two SE round, full post-selection; (b4) Different post-selection schemes.}
    \label{fig: logical op}
\end{figure*}

\subsection{Memory Experiments}
\label{sec:eval:memory}

Fig.~\ref{fig: memory} summarizes memory experiments across four axes.

\noindent\textbf{1. Surface code family (Fig.~\ref{fig: memory}a).}
We evaluate Z-basis memory for the rotated, unrotated, and toric surface codes at $d \in \{3, 5, 7\}$ under circuit-level noise.
All three variants exhibit increasing LER suppression as $d$ scales, aligned with theoretical results~\cite{shor1996fault, fowler2012surface}.
Notably, at fixed $d$, LER$_{\text{rotated}} >$ LER$_{\text{unrotated}} >$ LER$_{\text{toric}}$ under the threshold (0.8\%), reflecting the benefit of increasing qubit redundancy ($d^2$, $2d^2{-}2d{+}1$, and $2d^2$). The intuition is: when operating below the error threshold, more physical qubits provide more syndrome information that enables more accurate decoding, so the additional redundancy yields lower LER despite the larger number of error locations. 
Despite of having the same $d$, these codes show different LER performances, illustrating that \textit{code distance alone does not determine LER, and circuit-level simulation is essential for fair comparison}.

\vspace{2pt}
\noindent\textbf{2. BB code family (Fig.~\ref{fig: memory}b).} We reproduce the published BB code memory experiment~\cite{bravyi2024high}, matching per-round LER for all three instances within a factor of $2\times$, albeit statistical uncertainty and potentially different decoder settings. A distinct finding is that \frameworkname{}'s Pauli tracker automatically constructs additional detectors arising from the linearly dependent stabilizers intrinsic in BB codes, potentially beneficial for decoding (Table.~\ref{tab:correctness}).

\vspace{2pt}
\noindent\textbf{3. QEC efficiency (Fig.~\ref{fig: memory}c).}
Comparing LER per physical qubit at matched $p$, BB codes and 4D codes outperform the surface code family by 1--2 orders of magnitude, reflecting the advantage of qLDPC codes in QEC efficiency.
\frameworkname{} enables this cross-comparison within a single framework.

\vspace{2pt}
\noindent\textbf{4. SE circuit design (Fig.~\ref{fig: memory}d).}
We quantify the impact of SE circuit design by replacing the standard schedule~\cite{github_stim} with a variant that swaps the gate order of $Z$- and $X$-type stabilizers.
The swapped schedule is a valid SE circuit but loses the hook-error mitigation in the standard ordering~\cite{hirai2026no}, leading to more severe error propagation and a measurably higher LER.
\frameworkname{}'s modular design for QEC protocol implementation makes this ablation easy to do. 

\subsection{Logical Operations}
\label{sec:eval:logops}

\noindent\textbf{1. Transversal gates and lattice surgery (Fig.~\ref{fig: logical op}a).}
We benchmark transversal H, S, CNOT (TG) and LS-based CNOT against the memory baseline. All show error suppression scaling with $d$, confirming the correctness of \frameworkname{}, showing that it captures the spatial detectors required for correlated decoding of transversal gates, as well as the complex boundary merges and splits in lattice surgery. 
At $d$=$7$, $p$=$10^{-3}$, logical operations incur overhead over the memory baseline: $2.0\times$, $11.6\times$, $4.0\times$, and $7.9\times$ for H$_\text{TG}$, S$_\text{TG}$, CNOT$_\text{TG}$, and CNOT$_\text{LS}$, respectively, averaged over all sub-experiments (initial states and measurement bases). Notably, S$_\text{TG}$ exhibits much higher overhead than H$_\text{TG}$ and CNOT$_\text{TG}$: H$_\text{TG}$ contains transversal H and SWAP gates that merely relocate errors spatially, but S$_\text{TG}$ contains CZ gates that propagate a single-qubit $X$ error into a correlated two-qubit $X{\otimes}Z$ error within the same logical patch, producing higher-degree DEM hyperedges that compromise decoding accuracy.

\noindent\textbf{Key Insights.} Gate-specific overheads vary by an order of magnitude (2.0--11.6$\times$) and are not predictable from the memory baseline alone. Memory-baseline approximations~\cite{litinski2019game, gidney2025factor} remain useful for large-scale resource estimates, but our measurements show that gate-level corrections can meaningfully refine end-to-end fidelity predictions for near-term systems.

\noindent\textbf{2. State injection (Fig.~\ref{fig: logical op}b).}
The results show that LER scales linearly with $p$ and is independent of $d$ (Fig.~\ref{fig: logical op}(b1,b2)), as injection error is proportional to physical gate noise and occurs before QEC rounds can suppress it, consistent with prior work~\cite{yin2026iswitch, lao2022magic}. The $|Y\rangle$ state has higher LER than $|X\rangle$ and $|Z\rangle$ because it is sensitive to both $X$- and $Z$-type errors, whereas $|X\rangle$ ($|Z\rangle$) is corrupted only by $Z$ ($X$) errors. The corner injection protocol further exhibits a biased LER between $|Z\rangle$ and $|X\rangle$ (Fig.~\ref{fig: logical op}(b3)) due to asymmetric qubit initialization, while middle injection maintains balanced $|X\rangle/|Z\rangle$ LER from its symmetric initialization (Fig.~\ref{fig: logical op}(b1,b2)).

\noindent\textit{Post-selection trade-off.} \frameworkname{}'s pipeline allows reconfiguring which detectors trigger post-selection, exposing an explicit trade-off between the post-selection rate (PS-rate, proportion of retained samples) and LER. We compare three schemes on corner injection at $d{=}7$, $p{=}10^{-3}$ (Fig.~\ref{fig: logical op}(b4)): \emph{(1) Full PS} (post-select on all detectors) achieves the lowest LER at the cost of lowest PS-rate; \emph{(2) Full QEC} (no post-selection) retains all samples and induces higher LER; \emph{(3) Hybrid} (ours, post-select only detectors adjacent to the logical observable) improves PS-rate by $1.7\times$ over Full PS with only a $23\%$ LER increase, offering a favorable operating point.

\begin{figure*}[!ht]
    \centering
    \includegraphics[width=0.99\textwidth]{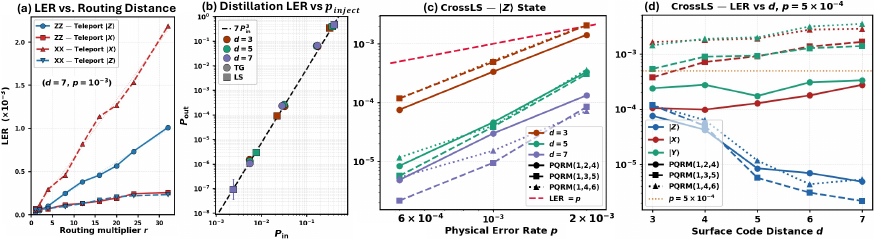}
    \caption{Surface-PQRM CrossLS protocol. (a) Routing Overhead in Bell state teleportation. (b) Distilled output fidelity matching theoretical $7p_{in}^3$. (c)(d) Surface-PQRM CrossLS protocol evaluation.}
    \label{fig: CrossLS}
\end{figure*}

\subsection{Logical Circuits}
\label{sec:eval:circuits}

\begin{figure}[!ht]
    \centering
    \vspace{-6pt}
    \includegraphics[width=0.48\textwidth]{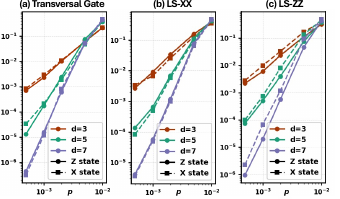}
    \caption{LER of Bell-state teleportation circuit implemented in (a) transversal gates and (b,c) lattice surgery.}
    \label{fig: logical circuit}
    \vspace{-6pt}
\end{figure}

\noindent\textbf{1. Bell-state teleportation.}
Fig.~\ref{fig: logical circuit} shows LER vs.\ $p$ for the three protocols (circuits in Fig.~\ref{fig: BellTeleport}), all exhibiting distance-dependent suppression. At $d=7$, $p=10^{-3}$, LERs are $3.0\times$, $9.4\times$, and $11.5\times$ above the memory baseline for TG, ZZ-LS, and XX-LS, respectively, quantifying the overhead from memory to logical circuits. We further vary inter-patch spacing to quantify LS routing overhead. Fig.~\ref{fig: CrossLS}(a) reveals a state-dependent asymmetry: teleporting $|Z\rangle$ ($|X\rangle$) via XX(ZZ)-LS incurs only sub-linear LER growth with routing distance, whereas teleporting $|Z\rangle$ ($|X\rangle$) via ZZ(XX)-LS yields strictly linear growth ($\mathrm{LER} = c \cdot r_{\rm inter} \cdot p^{\frac{d+1}{2}}$), increasing by ${\approx}\,3\times 10^{-5}$ per routing unit for ZZ-LS at $d=7$, $p=10^{-3}$ (${\approx}\,7\times 10^{-5}$ for XX-LS). The asymmetry stems from feed-forward correction structure: ZZ-LS teleportation of $|Z\rangle$ requires a correction $X^b$ whose fidelity depends on routing-spanning syndrome records from the $Z_1Z_2$ measurement, while $|X\rangle$ teleportation is invariant under $X^b$ correction and decouples from this accumulation (Fig.~\ref{fig: BellTeleport}(b)).

\noindent\textbf{Key insight.} The same logical function can exhibit drastically different LER depending on computing scheme, routing distance, and input state. These primitive-level overheads are only exposed by end-to-end simulation, and are exactly the building blocks needed for compositional FTQC evaluation. Detailed mechanism analysis in Appendix~\ref{app:bellderiv}.

\noindent\textbf{2. Distillation.}
Fig.~\ref{fig: CrossLS}(b) evaluates the Steane 7-to-1 distillation circuit (Sec.~\ref{sec:setup:existing}) under injection-only noise: only magic-state preparation is noisy while the distillation circuit itself is noiseless, matching the theoretical setting of noisy magic input with ideal Clifford gates~\cite{bravyi2005universal}. For both TG and LS, the output LER closely follows the theoretical $7p_{\rm in}^3$ curve~\cite{fowler2012surface, zhou2025low}, validating both circuit implementations and DEM construction.

\subsection{Surface-PQRM Lattice Surgery}
\label{sec:eval:crossls}

We present CrossLS as a case study demonstrating \frameworkname{}'s rapid prototyping capability for novel cross-code protocols. CrossLS teleports a logical state from a PQRM code to a surface code patch via lattice surgery, as described in Sec.~\ref{sec:setup:new}; we focus on the LER performance, while a full resource accounting vs.\ state-of-the-art magic state cultivation~\cite{gidney2024magic} is beyond the scope of this paper.

\noindent\textbf{LER vs.\ $p$.} Fig.~\ref{fig: CrossLS}(c) shows the $Z$-LER for all three PQRM instances across surface-code distances. At $d_{\text{surf}}=7$ and $p=10^{-3}$, PQRM(1,2,4) reaches ${\sim}3\times 10^{-5}$, 2 orders of magnitude below the PER. The intuition is: the $Z$-LER is governed by the logical $ZZ$ measurement during the merge phase, whose effective distance is $\min(d^X_{\text{surf}}, d^X_{\text{PQRM}})$ since logical $X$ error on either code will flip $ZZ$ measurement result. As both code grow, their $X$-distance grows, suppressing the $Z$-LER.

\noindent\textbf{LER vs.\ $d$ and state dependence.} Fixing $p=5\times10^{-4}$, we compare LER across logical states (Fig.~\ref{fig: CrossLS}(d)). For $|Z\rangle$, LER falls exponentially with $d_{\text{surf}}$ for all PQRM instances. For $|X\rangle$ and $|Y\rangle$, the trend reverses: larger PQRM instances fail to reach break-even, with LER increasing in $d_{\text{surf}}$; the smallest instance PQRM(1,2,4) $\llbracket 15,1,3 \rrbracket$ can break-even, but the LER suppression drops from $5\times$ below PER at $d=4$ to only $1.8\times$ at $d=7$. Structurally, these states are vulnerable to $Z$ errors, but this PQRM family has a fixed $d^Z_{\text{PQRM}}=3$ regardless of size. Increasing $d_{\text{surf}}$ therefore adds SE noise without adding $Z$-protection, worsening LER. $X$-stabilizer post-selection partially mitigates this, but a fundamental improvement would require scaling $d^Z_{\text{PQRM}}$, which in turn demands higher-weight $Z$-stabilizers that sacrifice the weight-4 local connectivity underlying CrossLS's hardware-friendliness.



%% file: TexFile/09_Related_Work.tex
\section{Related Work}
\label{sec:related}

\vspace{2pt}
\noindent\textbf{Clifford Simulation.}
Classical Clifford simulation rests on the Gottesman--Knill theorem~\cite{gottesman1997stabilizer} and its improved versions, from the Aaronson--Gottesman CHP algorithm~\cite{aaronson2004improved} to Stim~\cite{gidney2021stim}, which scales Pauli-frame simulation to billions of shots. They generate measurement samples consistent with a Clifford circuit's probability distribution, but remain agnostic to QEC protocol structure---detectors, logical observables, or protocol-level semantics are all outside their scope.

\vspace{2pt}
\noindent\textbf{Automated DEM Construction.} 
DEM construction therefore remains a manual, circuit-specific step on top of Stim. The spacetime codes formalism~\cite{delfosse2023spacetime, suau2026tqec} is the closest prior work toward automation, modeling QEC checks as codewords of a classical code induced by a physical Clifford circuit and identifying valid detectors. \frameworkname{} shares the underlying mathematical principle but operates at a different layer of the FTQC compilation stack: rather than analyzing a post-hoc physical circuit, \frameworkname{} \emph{bridges the logical and physical levels} by constructing the DEM progressively as the protocol is composed from atomic operations (Sec.~\ref{sec:tech2:ops}). This directly supports protocol-level semantics (e.g., lattice-surgery coupler transitions, feed-forward corrections) that have no clean representation at the physical-circuit level, and are therefore outside the reach of purely circuit-level tools. Beyond DEM construction, \frameworkname{} extends into a complete evaluation framework spanning QEC system representation, noise injection, and decoder backend.

\vspace{2pt}
\noindent\textbf{Protocol-Specific Implementations.}
Existing tools are protocol-specific and fragmented. For memory experiments, Stim provides built-in surface code circuits~\cite{github_stim} and Gong et~al.~\cite{github_sliding_window} provide BB code circuits. For logical operations, isolated Stim circuits exist for transversal gates~\cite{zhou2025low} and lattice surgery~\cite{stackexchange_lattice_surgery}. Bell-state teleportation has only a closed-source implementation~\cite{stack2025assessing}, while lattice-surgery distillation and cross-code protocols have no prior open-source implementations. FTPrimitiveBench~\cite{kan2026ftprimitivebench} benchmarks surface code primitives under structured noise, a slice of \frameworkname{}'s coverage across QEC codes and multi-patch protocols. End-to-end frameworks~\cite{litinski2019game, gidney2025factor, gidney2021factor} provide architectural resource estimates based on memory-baseline logical gate costs; \frameworkname{}'s gate-specific LER measurements (Sec.~\ref{sec:eval:logops}) supply the circuit-level data needed to refine such estimates.

%% file: TexFile/10_Conclusion.tex
\section{Conclusion}
\label{sec:conclusion}

We present \frameworkname{}, a framework that automates DEM compilation for the full spectrum of QEC protocols, eliminating the manual annotation bottleneck and unlocking systematic QEC evaluation beyond basic memory experiments. Correctness validation confirms exact detector and observable agreement with open-source references, and LER simulation results align with both references and theoretical predictions. Beyond validation, \frameworkname{} uncovers protocol behaviors and architectural insights inaccessible to memory-only simulation: gate-specific LERs exceed the memory baseline by up to $11.6\times$ (S$_\text{TG}$), LS routing distance contributes linearly to teleportation LER, and the CrossLS protocol exposes a state-dependent structural bottleneck invisible without end-to-end simulation. By unifying code families, protocols, and noise models in a single pipeline, \frameworkname{} serves as a foundational infrastructure for the rigorous evaluation and systematic exploration of future FTQC architectures.


%% file: TexFile/04_Motivation.tex
\section{End-to-End Examples for \frameworkname{}'s Pauli Tracker}
\label{sec:motivation}

We illustrate the core mechanics of \frameworkname{} using two concrete examples: two-qubit teleportation circuit realized in lattice surgery (LS) and transversal gates (TG), respectively. We demonstrate how tracking Pauli propagation systematically constructs DEMs based on two decomposition schemes: (1) for \textit{mid-circuit syndrome measurements}, current Paulis are decomposed into previously measured Paulis; and (2) for \textit{final data measurements}, the remaining system Paulis are decomposed into single-qubit data measurements.

\subsection{Example: Two-Qubit Teleportation}
\begin{figure}[!ht]
    \centering
    \includegraphics[width=0.4\textwidth]{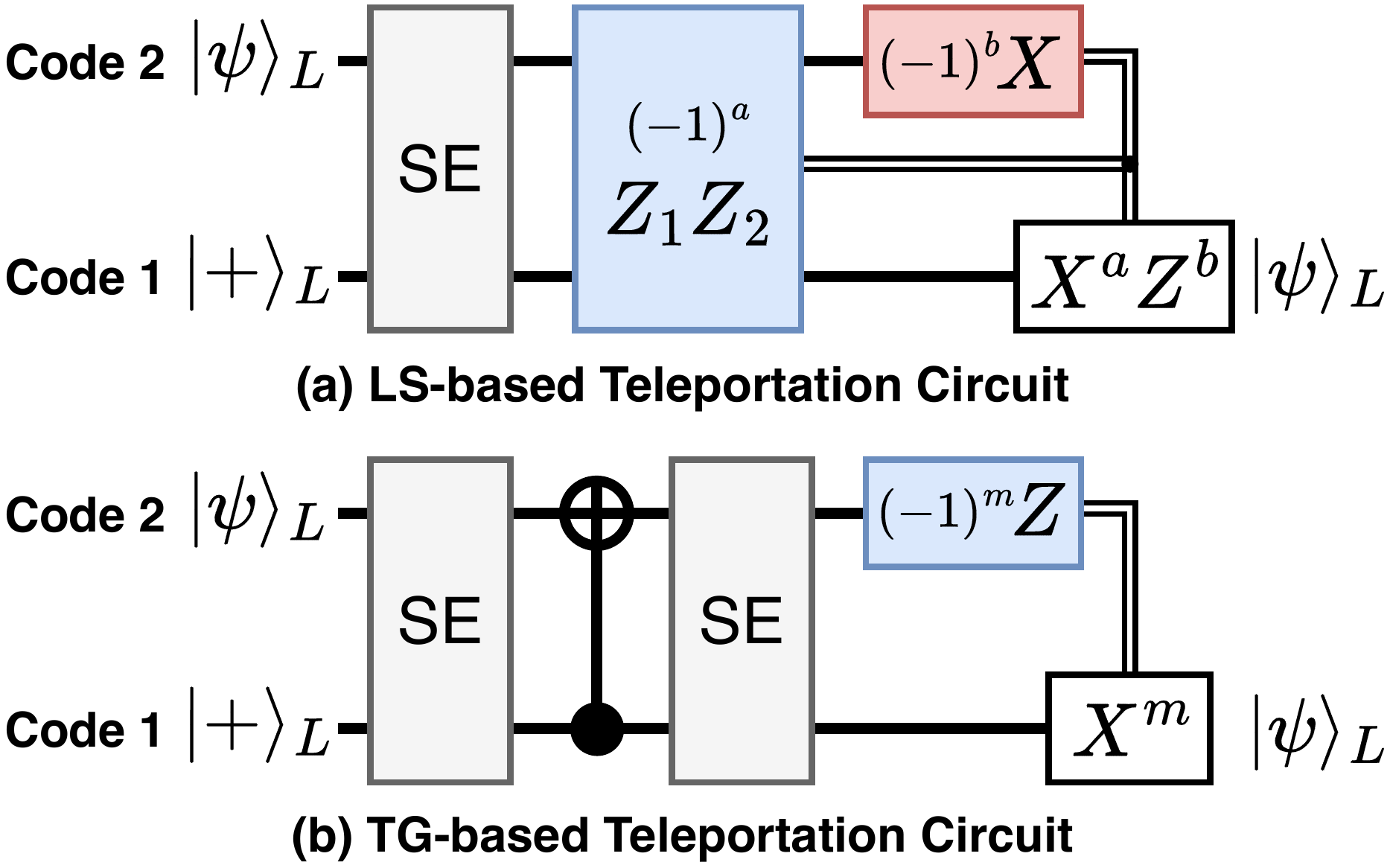}
    \caption{Two-qubit teleportation circuit realized in lattice surgery and transversal CNOT, respectively.}
    \label{fig: Teleportation}
\end{figure}

Fig.~\ref{fig: Teleportation} shows the teleportation circuit realized via LS and TG. Both initialize Code 1 in logical $|+\rangle_L$ and the target state $|\psi\rangle_L$ on Code 2, followed by an initial SE round. Their subsequent operations differ based on the underlying protocol. The LS circuit (Fig.~\ref{fig: Teleportation}(a)) entangles the patches via a joint logical $Z_1Z_2$ measurement (outcome $a$), measures Code 2 in the $X$-basis (outcome $b$), and applies a feed-forward Pauli correction ($X^a Z^b$) on Code 1. In comparison, the TG circuit (Fig.~\ref{fig: Teleportation}(b)) applies a transversal CNOT gate for entanglement, performs an intermediate SE round, and then measures Code 2 in the $Z$-basis (outcome $m$) to apply an $X^m$ correction. Although both circuits perform the exact same logical operation, their distinct physical structures lead to entirely different DEM constructions, which we detail below.

\begin{figure*}[!ht]
    \centering
    \includegraphics[width=0.99\textwidth]{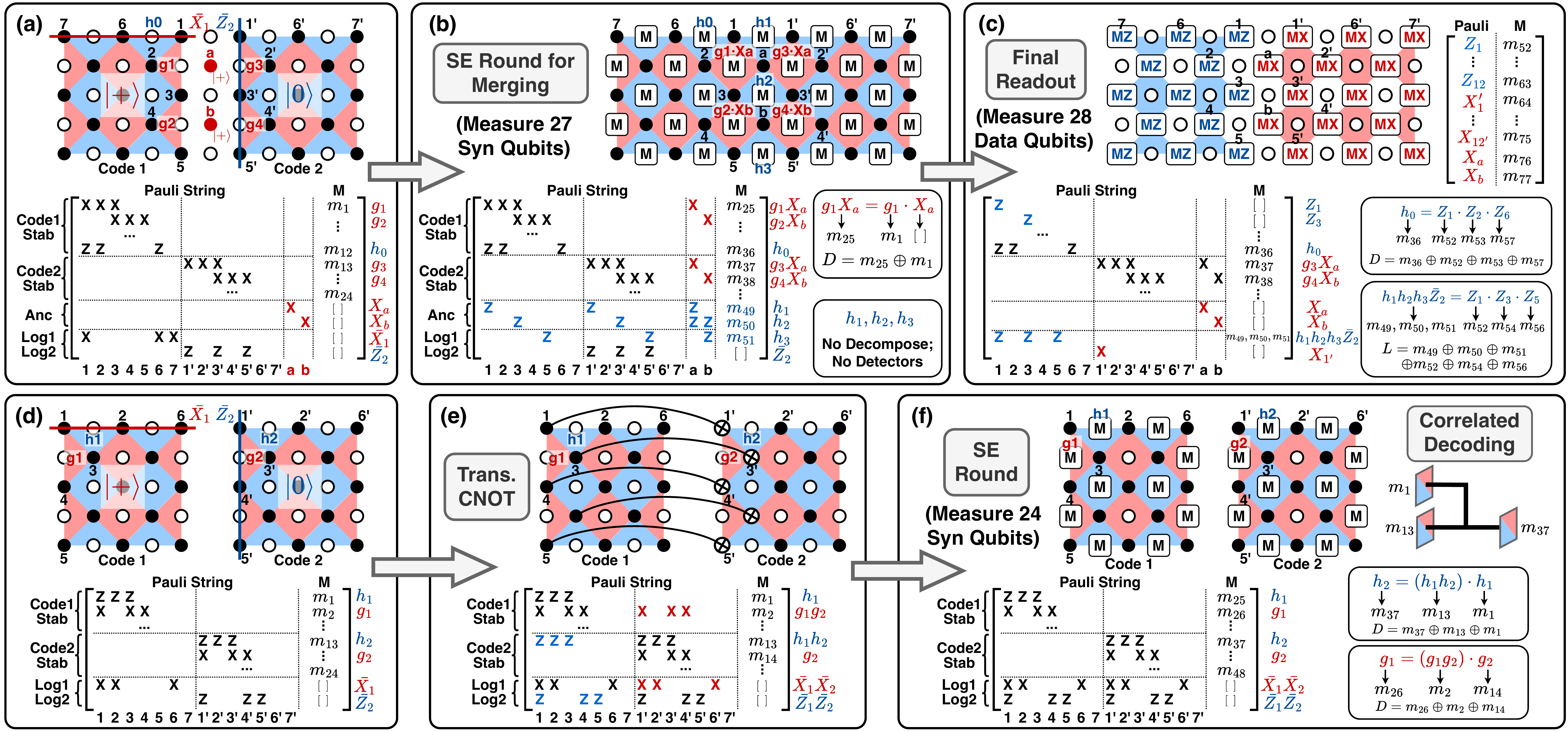}
    \caption{(a-c) Detector and logical observable construction in LS teleportation. (d-f) Detector construction in transversal CNOT gates, which naturally derives the correlated decoding technique.}
    \label{fig: MotivatingEx}
\end{figure*}

\subsection{LS-based Teleportation}
\label{sec:mot:LS}
Fig.~\ref{fig: MotivatingEx}(a-c) illustrates the physical execution and tableau evolution of the LS teleportation in Fig.~\ref{fig: Teleportation}(a), with $|\psi\rangle_L = |0\rangle_L$.

\noindent \textbf{(a) Initialization.} Following the initial SE round, the tableau tracks 24 stabilizers and their measurement records. The logical operators $\bar{X}_1 = X_1X_6X_7$ and $\bar{Z}_2 = Z_{1'}Z_{3'}Z_{5'}$ indicate that Code 1 and 2 are in $|+\rangle_L$ and $|0\rangle_L$, respectively. Two ancilla qubits ($a, b$) are initialized in $|+\rangle$, appending stabilizers $X_a, X_b$ (red) without measurement records to the tableau to mediate the upcoming patch merge.

\noindent \textbf{(b) Logical $Z_1Z_2$ Measurement.} The protocol measures the merged code stabilizers, introducing modified boundary $X$-stabilizers (e.g., $g_1X_a$) and 3 new $Z$-stabilizers ($h_1, h_2, h_3$), in total 27 measurements. It goes through two steps.

\noindent\textit{Tableau Update:} Because $h_1$ anti-commutes with $X_a$, $g_1$, and $g_3$, it replaces $X_a$ (the pivot) and multiplies into $g_1, g_3$, yielding $g_1X_a$ and $g_3X_a$ (following Sec.~\ref{sec:bg:tableau}). Similarly, $h_2$ updates $g_2, g_4$, and $h_3$ replaces the anti-commuting $\bar{X}_1$ (Fig.~\ref{fig: MotivatingEx}b), automatically capturing the loss of one logical degree of freedom as the two patches merge.

\noindent\textit{Detector Extraction:} We decompose current Paulis into previously measured ones. For example, $g_1X_a = g_1 \cdot X_a$. Summing their associated records yields a deterministic detector $D = m_{25} \oplus m_1$. Conversely, $h_1, h_2, h_3$ yield random outcomes upon anti-commuting replacement; thus, no valid decomposition or detector exists.

\noindent \textbf{(c) Final Readout.} Code 2 is measured in the $X$-basis to complete teleportation protocol (Fig.~\ref{fig: Teleportation}(a)), and Code 1 in the $Z$-basis for verification of teleported $|0\rangle_L$ from Code 2.

\noindent \textit{Tableau Update:} $X$-Paulis in Code 1 and $Z$-Paulis in Code 2 \& boundary are replaced by single-qubit $Z$ and $X$ Paulis, respectively, due to the anti-commuting measurement rule (Sec.~\ref{sec:bg:tableau}). Crucially, tracking the anti-commuting row multiplications yields a composite logical row $h_1h_2h_3\bar{Z}_2$ mapped to the accumulated records $m_{49}, m_{50}, m_{51}$.

\noindent\textit{Detector/Observable Extraction:} We decompose remaining Paulis into single-qubit data measurements (Fig.~\ref{fig: MotivatingEx}c, top-right). Decomposing the stabilizer $h_0$ yields a final-round detector involving both the syndrome record ($m_{36}$) and the data records ($m_{52},m_{53},m_{57}$). Meanwhile, decomposing the composite row $h_1h_2h_3\bar{Z}_2$ in the logical part of the tableau yields data measurements $m_{52}, m_{54}, m_{56}$. Summing all associated records forms the weight-6 logical observable $L$.

\vspace{2pt}
\noindent \textbf{Key Insight.} This algebraic derivation carries profound physical meaning: $m_{52}, m_{54}, m_{56}$ form the new logical $\bar{Z}_1$ on Code 1, while $m_{49}, m_{50}, m_{51}$ evaluate the logical $Z_1Z_2$ measurement outcome $a$ in Fig.~\ref{fig: Teleportation}(a). Their deterministic sum proves that the teleported state ($\bar{Z}_1$) after feed-forward correction ($X^a$) perfectly matches the original state ($|0\rangle_L$) on Code 2. \frameworkname{} natively and automatically tracks this end-to-end feed-forward propagation, which is a critical capability for compiling complex LS protocols.

\subsection{TG-based Teleportation}
\label{sec:mot:TG}

Fig.~\ref{fig: MotivatingEx}(d-f) illustrates the physical execution of the TG teleportation in Fig.~\ref{fig: Teleportation}(b), with $|\psi\rangle_L = |0\rangle_L$. We omit the final data measurement steps as their observable extraction is similar as that in Sec.~\ref{sec:mot:LS}.

\noindent\textbf{(d) Initial SE Round.} Similar to LS, the system initializes by measuring all stabilizers across both code patches with associated records.

\noindent\textbf{(e) Transversal CNOT.} Physical CNOT gates are applied pairwise between the data qubits of the two patches (Code 1 as control, Code 2 as target). Following Clifford update rules (Sec.~\ref{sec:bg:tableau}), $X$-Paulis propagate from control to target, and $Z$-Paulis propagate from target to control. Consequently, original stabilizers ($g_1, h_2$) and logicals ($\bar{X}_1, \bar{Z}_2$) evolve into $g_1g_2, h_1h_2, \bar{X}_1\bar{X}_2$, and $\bar{Z}_1\bar{Z}_2$, respectively (Fig.~\ref{fig: MotivatingEx}(e)).

\noindent\textbf{(f) Subsequent SE Round.} A second SE round measures the patches again. Applying our mid-circuit decomposition rule, these newly measured Paulis are decomposed into prior Paulis in the tableau. For instance, the current $h_2$ measurement ($m_{37}$) is decomposed into the previous $h_1h_2$ ($m_{13}$) and $h_1$ ($m_1$), yielding $h_2 = (h_1h_2)\cdot h_1$. Summing their associated records directly constructs the detector $D = m_{37} \oplus m_{13} \oplus m_1$ (Fig.~\ref{fig: MotivatingEx}(f), bottom-right). Similarly, $g_1$ is decomposed via $g_1 = (g_1g_2)\cdot g_2$ and yield its corresponding detector.

\vspace{2pt}
\noindent \textbf{Key Insight: Automated Correlated Decoding.} Notably, the derived detector ($D = m_{37} \oplus m_{13} \oplus m_1$) spans one current record and two previous records across \textit{both} spatial patches (Fig.~\ref{fig: MotivatingEx}(f), top-right). This cross-patch structure is exactly the \textit{correlated decoding} technique recently proposed to handle transversal gate errors~\cite{cain2024correlated, zhou2025low}. While manual derivation of such detectors becomes intractable for complex, multi-patch transversal operations, \frameworkname{}'s systematic Pauli tracking inherently and robustly automates this construction.

%% file: TexFile/11_AppendixB.tex
\section{Bell Teleportation Routing-Distance Mechanism}
\label{app:bellderiv}

\begin{figure*}[!ht]
    \centering
    \includegraphics[width=0.99\textwidth]{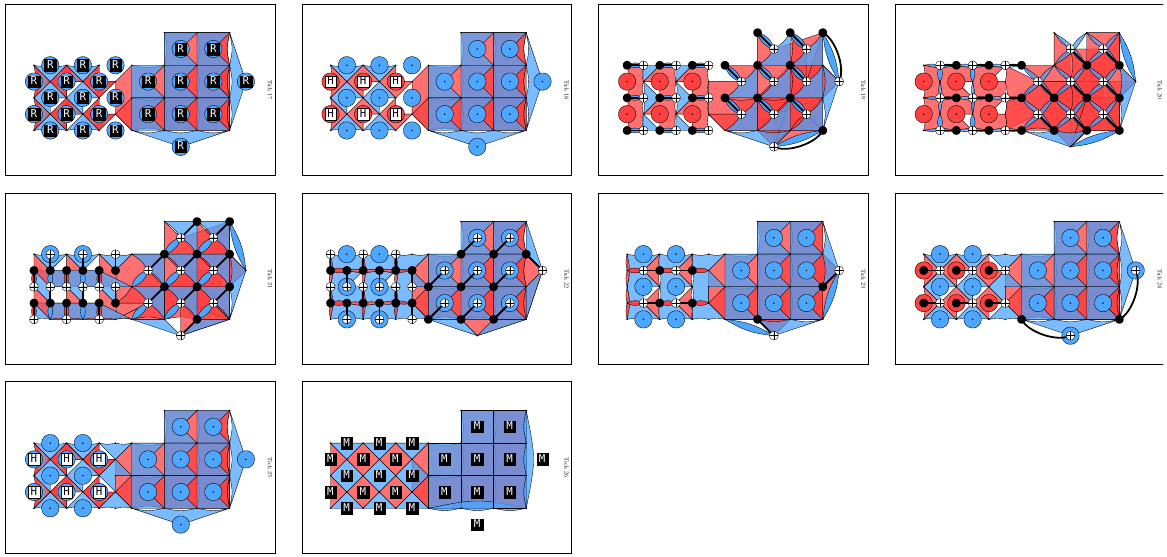}
    \caption{The tick-accurate cross-code SE circuit scheduling for CrossLS protocol.}
    \label{fig: CrossLS scheduling}
\end{figure*}

This appendix derives the linear LER accumulation with routing distance for $ZZ$-LS Bell teleportation (Sec.~\ref{sec:eval:circuits}).

Consider the ZZ-LS circuit (Fig.~\ref{fig: BellTeleport}b) teleporting $|Z\rangle_L$ from Code~2 to Code~3. The lattice-surgery merge measures $Z_1Z_2$ over $r_{\rm inter}$ routing columns connecting the two patches. The merge outcome $b$ is the XOR of syndrome records from every ancilla qubit spanning the interspace. Any $X$-type error occurring in the interspace during the merge rounds flips one of these syndrome records and hence flips $b$, corrupting the feed-forward correction $X^b$ applied to Code~3.

Formally, let $r_{\rm inter}$ denote the number of qubit columns in the routing region. Each column contributes $d \cdot (d-1)/2$ ancilla measurements per SE round, each susceptible to $X$-type errors at rate $\sim p$. The number of error-sensitive locations scales as $d \cdot r_{\rm inter}$, giving a logical failure probability
\begin{equation}
    \mathrm{LER}(r_{\rm inter}) \approx c \cdot r_{\rm inter} \cdot p^{\frac{d+1}{2}},
\end{equation}

Here, $c$ absorbs constant prefactors. This linear dependence on $r_{\rm inter}$ is confirmed experimentally (Fig.~\ref{fig: logical circuit}a2): ZZ-LS teleporting $|Z\rangle$ increases LER by $\approx 3\times 10^{-5}$ per routing unit at $d=7$, $p=10^{-3}$.

For $|X\rangle$ teleportation via ZZ-LS, the feed-forward correction is still $X^b$, but applying $X$ to an $X$-eigenstate leaves it invariant. The $Z_1Z_2$ outcome $b$ is therefore irrelevant to LER, decoupling the routing-distance effect entirely. The same argument applies symmetrically to XX-LS: XX-LS teleporting $|X\rangle$ accumulates errors linearly, while XX-LS teleporting $|Z\rangle$ does not.

\frameworkname{}'s Pauli tracker automatically decomposes the $Z_1Z_2$ outcome $b$ into its constituent syndrome records at compile time, directly revealing this accumulation structure without any manual analysis. The routing-distance asymmetry is thus a free byproduct of automated DEM construction.

%% file: TexFile/12_AppendixC.tex
\section{SE Circuit Scheduling for CrossLS}
\label{sec: tick scheduling}
Fig.~\ref{fig: CrossLS scheduling} presents the SE circuit scheduling for the CrossLS protocol. The $Z$-stabilizers of the PQRM code, the boundary $Z$-stabilizers, and all stabilizers of the surface code, are measured simultaneously. The CNOT gates span 6-ticks. Only the two boundary $Z$-stabilizers require long-range connectivity, while the rest remain local.